\documentclass[aps,prf,longbibliography,onecolumn]{revtex4-2}  
\usepackage{graphicx,color,subfigure}
\usepackage{amsmath}

\usepackage{subfigure}
\usepackage{hyperref}

\newcommand\be{{\mbox{\boldmath $e$}}}
\newcommand\bu{{\mbox{\boldmath $u$}}}
\newcommand\bg{{\mbox{\boldmath $g$}}}
\newcommand\bJ{{\mbox{\boldmath $j$}}}
\newcommand\bj{{\mbox{\boldmath $j$}}}

\newcommand\bB{{\mbox{\boldmath $B$}}}

\newcommand\vor{\boldsymbol{\omega}}
\newcommand \del{\nabla}

\newcommand\Ra{\mbox{Ra}}

\newcommand\Pra{\mbox{Pr}}
\newcommand\Pm{\mbox{Pm}}
\newcommand\Tay{\mbox{Ta}}
\newcommand\Ek{\mbox{Ek}}

\newcommand \Tpert{\theta}
\newcommand\eg{e.g.\ }
\newcommand\ie{i.e.\ }
\newcommand\emf{\mathcal{E}}
\newcommand\bemf{\boldsymbol{\mathcal{E}}}
\newcommand\Rm{\mbox{Rm}}
\newcommand\lb{\ell_b}
\newcommand\lu{\ell_u}
\newcommand\Urms{U_{\ast}}
\newcommand\Brms{B_{\ast}}
\newcommand\Bmrms{\overline{B}_{\ast}}

\begin{document}
\title{Subcritical dynamos in rapidly-rotating planar convection}
\author{R.~G. Cooper}
\email{r.cooper4@newcastle.ac.uk}
\author{P.~J. Bushby}
\author{C. Guervilly}
\affiliation{School of Mathematics, Statistics and Physics,
Newcastle University, Newcastle upon Tyne NE1 7RU, UK}

\begin{abstract}
We study dynamo action using numerical simulations of planar Boussinesq convection at rapid rotation (low Ekman numbers, $\Ek$), focusing on subcritical dynamo action in which the dynamo is sustained for Rayleigh numbers, $\Ra$, below the critical Rayleigh number for the onset of non-magnetic convection, $\Ra_c$. These solutions are found by first investigating the supercritical regime, in which the dynamo is able to generate a large-scale magnetic field that significantly influences the convective motions, with an associated Elsasser number of order $\Ek^{1/3}$. Subcritical solutions are then found by tracking this solution branch into the subcritical regime, taking a supercritical solution and then gradually lowering the corresponding Rayleigh number. We show that decreasing the Ekman number leads to an extension of the subcritical range of $\Ra/\Ra_c$, down to an optimal value of $\Ek=10^{-5}$. For magnetic Prandtl numbers of order unity, subcriticality is then hampered by the emergence of a large-scale mode at lower Ekman numbers when the dynamo driven by the smaller scale convection generates relatively stronger large-scale magnetic field. The inability of the large-scale mode to sustain dynamo action leads to an intermittent behaviour that appears to inhibit subcriticality. The subcritical solutions are also sensitive to the value of the magnetic Reynolds number (or equivalently, the magnetic Prandtl number, $\Pm$), as values of the magnetic Reynolds number greater than $70$ are required to produce dynamo action, but large values lead to fluctuations that are able to push the system too far from the subcritical branch and towards the trivial conducting state. 
\end{abstract}

\maketitle


\section{Introduction}
\label{intro}

Understanding the physical mechanisms that generate magnetic fields in electrically-conducting fluids is one of the fundamental challenges in magnetohydrodynamics. 
In most astrophysical bodies, magnetic fields are thought to be generated by a hydromagnetic dynamo, which converts the kinetic energy of fluid motions into magnetic energy, leading to the amplification of the magnetic field.
In the case of the geodynamo for example, convectively-driven motions in the liquid core are responsible for the generation of the Earth's magnetic field. 
Whilst numerical simulations have successfully replicated many of the features of the geodynamo  \citep[e.g.][]{Gla95,Christensen2010,Christensen2015TOG}, it is currently not possible to carry out simulations at Earth-like parameter values. 
One particularly problematic parameter for numerical simulations is the Ekman number, $\Ek$, which estimates the strength of the viscous forces with respect to the strength of the Coriolis force.
With considerable computational effort, modern spherical geodynamo calculations can reach Ekman numbers as small as $\Ek=10^{-7}$ \citep{Sch17,Aubert2019}, but this is still approximately 8 orders of magnitude larger than the estimated Ekman number for the Earth's liquid core, so all such simulations overestimate the relative importance of viscous forces compared to the Coriolis force. 
Whilst it may be possible to use scaling laws \citep[e.g.][]{Chr06} to bridge the gap in parameter space between numerical dynamos and planetary dynamos, this remains controversial \citep{Stelzer2013,KB_2013,Oruba2014}, thus further exploration of the low Ekman number regime is necessary. One productive approach to this problem is to consider the more idealised case of convectively-driven dynamos in simpler geometries, for example a plane layer.  
Although there are still limitations on how small the Ekman number can be made, these simpler problems are more amenable to a systematic study of the computationally accessible regions of parameter space. 

Many previous studies have focused upon layers of electrically-conducting fluid in uniformly rotating Cartesian domains, in which convection is driven by heating the layer from below. 
If the layer is rotating about the vertical axis (with the angular velocity vector anti-parallel to the constant gravitational acceleration) then it is well known that rotation tends to inhibit convection; furthermore, near convective onset, the horizontal lengthscale of the convective motions decreases with increasing rotation rate, leading to tall, narrow convective cells \cite{Chandra53}. 
For a rapidly-rotating domain, Childress \& Soward \citep{Child72,Sow74} were the first to demonstrate that these near-onset convective motions are capable of sustaining a dynamo. 
One of the key results of their study was the identification of several possible force balances in the governing equations, depending upon the strength of the generated magnetic field.
These balances suggest the existence of multiple dynamo branches.  
If the induced magnetic field is weak \citep{Sow74}, it plays a negligible dynamical role and the dominant force balance is geostrophic.
In the (so-called) ``intermediate'' and ``strong'' field limits \citep[e.g.][]{Fautrelle1982}, the Lorentz force plays an increasingly important role in the dynamics of the dynamo, significantly perturbing the driving flow.

Different aspects of the Childress-Soward dynamo have been explored by many authors \citep[e.g.][]{StP93,JonesRoberts2000,RotvigJones2002,Stel04,Favier2013b,Calkins2015b}. 
Arguably the most relevant study to that of the present paper is that of \citet{Stel04}, who carried out a range of convective dynamo simulations at low $\Ek$ (down to $\Ek=5\times10^{-7}$). 
The level of convective driving in such systems is often quantified in terms of the Rayleigh number, $\Ra$, which is a dimensionless measure of buoyancy effects relative to diffusion. 
\citeauthor{Stel04} focused primarily upon a weakly supercritical regime in which the Rayleigh number was only 15-20$\%$ above its critical value at the non-magnetic convective onset under the influence of rotation (denoted $\Ra_c$). 
They found efficient dynamos, characterised by an initial phase of exponential growth in the magnetic energy, followed by a brief period of super-exponential growth as the magnetic field becomes strong enough to influence the flow (also accompanied by a similar growth in the kinetic energy). The dynamo then saturates, reaching a statistically steady state in which the magnetic energy typically exceeds the kinetic energy of the system. 
We can gain some insight into the mechanisms responsible for this phase of super-exponential growth by considering related magnetoconvection studies \citep{Chandra61,Elt72}. Like rotation, imposed magnetic fields tend to have a stabilising effect upon convection in electrically-conducting fluids, raising the corresponding critical Rayleigh number for the onset of convection. 
However, when magnetic fields are applied across rapidly-rotating domains, the Lorentz force can relax some of the rotational constraints, thus facilitating convection rather than impeding it (of course, this depends crucially upon the field strength; the Lorentz force and the Coriolis force must be of comparable magnitudes at a certain lengthscale).
In these dynamo calculations, phases of super-exponential growth are indicative of a similar effect, with the dynamo-generated magnetic field becoming locally strong enough to break some of the rotational constraints.     
The field that is produced significantly perturbs the flow, allowing the dynamo to reach a balanced state in which the magnetic field plays an important dynamical role in the nonlinear dynamo.
Low Ekman number dynamos that are able to sustain sufficiently strong magnetic fields that play an important dynamical role are relevant for astrophysical systems such as the geodynamo \citep{Gil10}, but remain largely unexplored due to the difficulty in performing numerical simulations in this regime. The case described in \citeauthor{Stel04} is an important example of such a dynamo, which we explore further in the present paper.

One of the most interesting findings of \citeauthor{Stel04} (see also the related study of St Pierre \cite{StP93}) was the fact that it is possible to follow dynamo solution branches into the subcritical regime, which is characterised by a Rayleigh number that is less than $\Ra_c$. This branch tracking is achieved by taking a supercritical dynamo solution as an initial condition and then gradually lowering the Rayleigh number. 
The existence of this subcritical behaviour is a direct consequence of the field having changed the flow geometry. 
If the magnetic field were to be removed, the system would gradually relax back to a hydrostatic state; to put this another way, these subcritical solutions exhibit bistability with the trivial conducting state.
 Examples of subcritical dynamo action have also been found in rotating spherical shells \cite{Mor09,Sree11,Dor16}, but there the subcriticality is less extreme: nonlinear dynamo solutions can be found below the critical Rayleigh number for the onset of dynamo action, but not below the critical Rayleigh number for the onset of convection. 
It is worth emphasising that the existence of subcritical dynamos is more than simply a mathematical curiosity: it has been suggested that the comparatively rapid cessation of the Martian dynamo could have been a direct consequence of the Martian dynamo reaching the end of a subcritical dynamo branch \cite{Kua08,Hor13}.

There has not yet been a systematic study of the parametric dependence of subcritical convectively-driven dynamos in rapidly rotating Cartesian domains. Whilst individual examples of such subcritical dynamos have been presented \cite{StP93,Stel04}, it is not clear what the optimal parameter regime is for the existence of these solutions. The primary aim of this paper is to address this issue. In the next section, we begin by presenting the governing equations and defining the output quantities used for the analysis of the data. Section~\ref{sec:Results} presents the results of our study, beginning with an analysis of near-onset supercritical dynamo action before decreasing the Rayleigh number below the linear onset of convection to explore subcritical dynamos. Finally, we summarise our key findings in Section~\ref{sec:discussion}.

\section{Model}


\subsection{Governing equations and boundary conditions}

We study dynamo action driven by rotating Boussinesq convection in a three-dimensional planar model. The computational domain has depth $d$ in the vertical direction and width $\lambda d$ in each of the horizontal directions.  The gravitational acceleration is $\bg=-g \be_z$, where $\be_z$ is the unit vector in the $z$-direction. The layer rotates about the vertical axis at constant rotation rate $\Omega$. A vertical temperature difference $\Delta T$ is imposed across the layer such that the fluid is uniformly heated from below. The fluid has constant kinematic viscosity $\nu$, thermal diffusivity $\kappa$, magnetic diffusivity $\eta$, density $\rho$, thermal expansion coefficient $\alpha$ and magnetic permeability $\mu_0$. The system of governing equations is solved in dimensionless form. We scale lengths with $d$, times with $d^2/\kappa$ (the thermal diffusion timescale), velocities with $\kappa/d$, temperature with $\Delta T$ and magnetic field with  $(\mu_0 \rho)^{1/2}\kappa/d$. The non-dimensional governing equations are
\begin{eqnarray}
&& \frac{\partial \bu}{\partial t} + \bu \cdot \del \bu + \frac{\Pra}{\Ek} \be_z \times \bu  
=  -\del p + \Pra \del^2 \bu + \Pra \Ra \Tpert \be_z + \bJ \times \bB,
\\ 
&&\del \cdot \bu = 0,
\\ 
&& \frac{\partial \Tpert}{\partial t} + \bu \cdot \del \Tpert = u_z + \del^2 \Tpert,
\\
&& \frac{\partial \bB}{\partial t} + \bu \cdot \del \bB - \bB \cdot \del \bu = \frac{\Pra}{\Pm}\del^2 \bB,
\\
&&\del \cdot \bB = 0,
\end{eqnarray}
where $\bu$ is the velocity, $p$ the pressure, $\bB$ the magnetic field, $\Tpert$ the temperature fluctuation from the linear background profile, and $\bJ = \del \times \bB$ the current density.
The dimensionless parameters that appear in the governing equations are the Prandtl number $\Pra$, magnetic Prandtl number $\Pm$, Ekman number $\Ek$, and Rayleigh number $\Ra$, which are defined as
\begin{equation}
\Pra = \frac{\nu}{\kappa}, \quad \Pm = \frac{\nu}{\eta}, \quad \Ek = \frac{\nu}{2\Omega d^2}, \quad   \Ra=\frac{\alpha g \Delta T d^3}{\kappa \nu}.
\end{equation} 
In order to avoid the numerical challenges of solving viscous boundary layers in the low Ekman number regime we choose the lower and upper boundaries to be stress-free. Furthermore this allows us to compare our results directly with those of previous studies such as \citet{Stel04}. The lower and upper boundaries are also impermeable and perfect thermal and electrical conductors, i.e.
\begin{equation}
	\Tpert = u_z = \frac{\partial u_x}{\partial z} = \frac{\partial u_y}{\partial z} = 
	B_z = \frac{\partial B_x}{\partial z} = \frac{\partial B_y}{\partial z} = 0 \quad \textrm{at} \quad z=\left\{0,1\right\},
	\label{eq:BC}
\end{equation}
with periodic boundaries in the $x$ and $y$ directions. Simulations are initiated by a weak small-scale temperature pertubation to the trivial conducting state in the presence of a weak imposed small-scale horizontal magnetic field with zero net flux.

The governing equations are solved using the pseudospectral code described in \citet{Cat03}, where more details about the numerical scheme can be found.


\subsection{Definition of the output quantities}
\label{sec:def}

The magnetic Reynolds number, $\Rm$, gives an estimate of the ratio of magnetic induction to magnetic diffusion. $\Rm$ is based on the layer depth and the root mean square (r.m.s) velocity amplitude. In terms of our dimensionless variables,
\begin{equation}
	\Rm = \frac{\Pm}{\Pr} \Urms,
\end{equation} 
where $\Urms=\sqrt{\langle \bu^2 \rangle_V}$. Throughout this paper we denote volume averages as $\langle (...) \rangle_V$ with $V$ the volume of the computational domain, and horizontal averages over $x$ and $y$ as $\overline{(...)}$. The ratio of magnetic to kinetic energy is 
\begin{equation}
 M=\frac{\Brms^2}{\Urms^2},
\end{equation} 
with $\Brms=\sqrt{\langle \bB^2 \rangle_V}$.

\par As previously noted, one of the key balances in this system is the comparison between Coriolis effects and the Lorentz force in order to determine the influence of the magnetic field on the dynamics of the system. In particular we are interested in the ability of the magnetic field to loosen some of the rotational constraints which inhibit convection. Motivated by these considerations, we therefore use the Elsasser number, $\Lambda$, which measures the ratio of the Lorentz force to Coriolis forces, as the primary measure of the magnetic field strength. In these dimensionless units,

\begin{equation}
	\Lambda=\frac{\Pm \Ek}{\Pra^2} \Brms^2.
\end{equation}
Given that these dynamos produce a non-negligible mean magnetic field (in a horizontally-averaged sense), it is also of interest to consider an alternative definition of the Elsasser number that is based upon this mean field,

\begin{equation}
	\overline{\Lambda}=\frac{\Pm\Ek}{\Pra^2}\Bmrms^2,
\end{equation}
where $\Bmrms=\sqrt{\langle \overline{\bB}^2 \rangle_{V}}$ is the r.m.s. amplitude of the horizontally-averaged magnetic field. These definitions of the Elsasser number are based on the standard formulation for this quantity, which assumes (in dimensional terms) that $d$ is a possible appropriate lengthscale for the dynamo and that $\eta/d$ 
is an appropriate velocity scale (equivalently this assumes that the appropriate timescale is ohmic, based on the layer depth). Another important lengthscale for the magnetic field is the magnetic dissipation lengthscale, $\lb$, defined by 

\begin{equation}
	\lb^2=\frac{\langle \bB^2 \rangle_{V}}{\langle |\del \times \bB|^2 \rangle_{V}}.
	\label{eq:lb}
\end{equation}

Having introduced appropriate lengthscales for the magnetic field, it is natural to do the same for the flow. Recalling that the convective cells near onset are long and thin, the vertical scale is roughly unity in these units. Our estimate for the typical flow lengthscale in the horizontal direction is based on the vertical velocity,
\begin{equation}
	\lu = \frac{\sum\limits_{k_x,\, k_y,\, k_z} \left(\hat{u}_z(k_x, k_y, k_z) \right)^2}
	{ \sum\limits_{k_x,\, k_y,\, k_z}  \sqrt{k_x^2+k_y^2} \left(\hat{u}_z(k_x, k_y, k_z)\right)^2},
	\label{eq:lu}	   
\end{equation}
where $\hat{u}_z$ denotes the three-dimensional Fourier transform of the vertical velocity, whilst $k_x$, $k_y$ and $k_z$ are the wavenumbers in the $x$, $y$ and $z$ directions respectively \citep{Guer14}. A similar estimate can be made for the vertical flow lengthscale, replacing $\sqrt{k_x^2+k_y^2}$ with $k_z$.

The efficiency of convective heat transport produced by the flow is quantified via the ratio of total heat flux to conductive heat flux in the absence of motions, known as the Nusselt number,  
	\begin{equation}
	\mbox{Nu}=1+\langle |\del \theta|^2 \rangle_V.
	\end{equation} 

Arguably the most natural timescale to use for dynamo calculations is the magnetic diffusion (ohmic) timescale, $t_{\eta}$, based on the depth of the domain. In dimensional terms, $t_{\eta}=d^2/\eta$, which is a factor of $\Pm/\Pr$ longer than our characteristic timescale (based on the thermal diffusion time). This estimate for the relevant dynamo timescale is likely to be an overestimate, as it is based on the molecular magnetic diffusivity as opposed to the turbulent magnetic diffusivity. Following \citet{Cat91b}, we define an alternative characteristic timescale, $t_e$, which is an estimate for the decay time of the large-scale field due to turbulent diffusion. In dimensional terms,
\begin{equation}
	t_e = \left(\frac{d}{\lu}\right)^2 \frac{\langle \bB^2 \rangle_{V}}{\eta \langle |\del \times \bB|^2 \rangle_{V}}.
\end{equation} 
In dimensionless form, $t_e=(\Pm/\Pr)(\lb^2/\lu^2)$. \citet{Cat91b} argue that a dynamo exists only if the field survives for times much longer than $t_e$; all of the successful dynamo calculations that are reported in this paper satisfy this condition.

Finally, in dynamos driven by spherical rotating convection, \citet{Sree11} show that dipolar magnetic fields lead to an increase in helicity due to feedback on the flow. This enhancement in the helicity then leads to subcritical behaviour in their simulations. In this Cartesian domain, the kinetic helicity at each depth is given by
\begin{equation}
\mathcal{H}=\overline{\bu \cdot \del \times \bu}.
\end{equation}
The relative kinetic helicity normalises the helicity relative to the flow intensity. At each depth this is then given by
\begin{equation}
\mathcal{H}_{rel}=\frac{\overline{\bu \cdot \del \times \bu}}{\left(\overline{\bu^2}\right)^{1/2} \left(\overline{\left(\del \times \bu \right)^2}\right)^{1/2}},
\end{equation}
where $|\mathcal{H}_{rel}|\leq 1$. Maximally helical flows produce a relative kinetic helicity of $\mathcal{H}_{rel}=\pm 1$.


\section{Results}
\label{sec:Results}
Table~\ref{table1} summarises the input and output quantities of the numerical simulations reported in this paper. The Ekman number is varied in the range $\Ek\in[5\times10^{-7},3.16\times10^{-4}]$ and the magnetic Prandtl number is in the range $\Pm\in[1,5]$.  In rotating convection studies, the Taylor number $\Tay=1/\Ek^2$ is often used as an alternative to the Ekman number; our upper limit of $\Ek=3.16\times10^{-4}$ corresponds to $\Tay=1\times10^7$. We focus on dynamos generated near the onset of convection and the Rayleigh number is varied between  $0.85\Ra_c$ and $1.18\Ra_c$, where $\Ra_c$ is the critical Rayleigh number at the linear onset of rotating convection calculated from the analytical study of \citet{Chandra61}. The standard Prandtl number is fixed to be $\Pra=1$ in all cases. 

\begin{figure}
\centering
\includegraphics[width=0.6\textwidth]{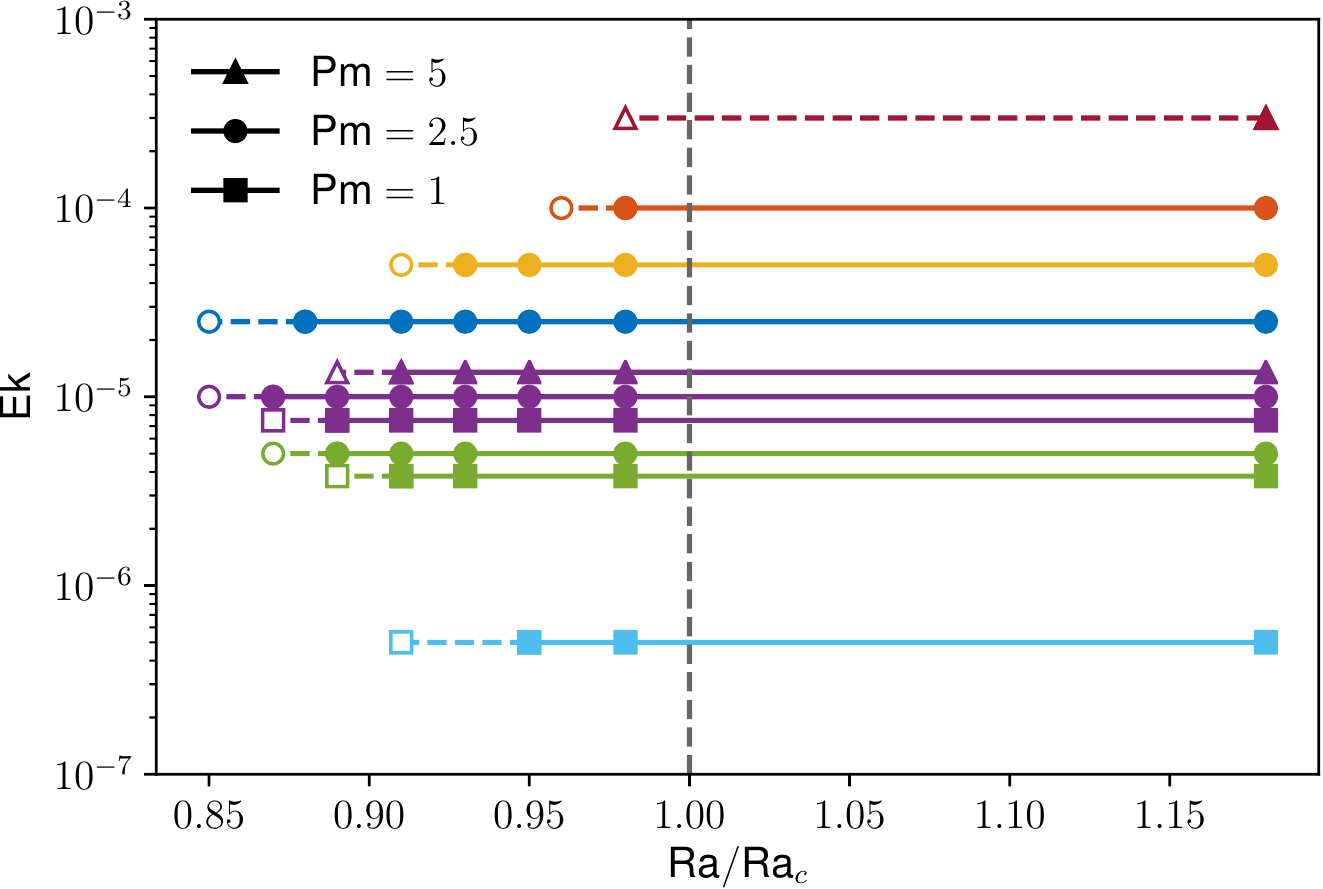}
\caption{Location of the simulations in the parameter space $\Ek$-$\Ra/\Ra_c$. Filled symbols represent cases where a dynamo is sustained and the open symbols represent cases where the dynamo fails. The marker colour denotes $\Ek$ and the marker shape $\Pm$. For $\Ek=10^{-5}$ and $\Ek=5\times 10^{-6}$, points with different $\Pm$ have been shifted vertically for visibility. It should be noted that an infinite time series would be required to conclusively state the existence of a subcritical dynamo. The majority of the cases presented here persist for long integrations (for a period of time much greater than $t_e$), however cases at $\Ek=5\times 10^{-7}$ were only run for a comparatively short time (see Table~\ref{table1}).
}
\label{fig:runs}
\end{figure}

Figure~\ref{fig:runs} shows the location of the dynamo simulations in the parameter space $\Ek$-$\Ra/\Ra_c$, where simulations with different magnetic Prandtl numbers are indicated by different marker shapes. The behaviour of the dynamo changes drastically at the lowest Ekman number that we calculated ($\Ek=5\times10^{-7}$), so the regimes with moderate Ekman numbers where $\Ek\geq5\times10^{-6}$ (\S\ref{sec:super} and \ref{sec:sub}) and with small Ekman numbers where $\Ek=5\times10^{-7}$(\S\ref{sec:E5e-7}) are presented separately.
To benchmark some of our results, we have reproduced the dynamo results of \citet{Stel04} (hereafter SH04). The benchmarked cases are indicated by an asterisk in Table~\ref{table1}. Values of $\Rm$ and $\Lambda$ are in agreement with the results of SH04 within 5\%. 

\begin{table*}
\begin{center}
\scriptsize
\begin{tabular}{c c c c c c c c c c c c c c c}
Case & $\Ek$ & $\Ra$ & $\Ra/\Ra_c$ & $\Pm$ & $\lambda$ & $N_x\times N_y\times N_z$ & $t_r$ & $t_r/t_e$ & $\Rm$ & $\ell_b$ & $\ell_u$ & $\Lambda$ & $\overline{\Lambda}$ & $M$\\
\hline
A1 & $3.16\times 10^{-4}$ & $4.8934\times 10^5$ & $1.18$ & $5$ & $2$ & $128^3$ & $17.28$ & $198.0$ & $123\pm11$ & $0.044$ & $0.33$ & $0.30\pm0.13$ & $0.030$ & $0.31$ \\
A2 & $3.16\times 10^{-4}$ & $4.0640\times 10^5$ & $0.98$ & $5$ & $2$ & $128^3$ & $1.30$ & $-$ & $-$ & $-$ & $-$ & $-$ & $-$ & $-$\\
\hline
B1* & $1 \times 10^{-4}$ & $2.2345 \times 10^{6}$ & $1.18$ & $2.5$ & $1$ & $128^3$ & $9.84$ & $133.2$ & $108\pm18$ & $0.041$ & $0.24$ & $0.30\pm0.14$ & $0.061$ & $0.63$ \\
B2 & $1 \times 10^{-4}$ & $1.8619 \times 10^{6}$ & $0.98$ & $2.5$ & $1$ & $128^3$ & $6.56$ & $72.4$ & $71\pm15$ & $0.048$ & $0.25$ & $0.26\pm0.13$ & $0.079$ & $1.25$ \\
B3 & $1 \times 10^{-4}$ & $1.8201 \times 10^{6}$ & $0.96$ & $2.5$ & $1$ & $128^3$ & $1.34$ & $-$ & $-$ & $-$ & $-$ & $-$ & $-$ & $-$ \\
\hline
C1* & $5 \times 10^{-5}$ & $5.6050 \times 10^6$ & $1.18$ & $2.5$ & $1$ & $128^3$ & $4.51$ & $58.4$ & $141\pm24$ & $0.035$ & $0.20$ & $0.36\pm0.11$ & $0.068$ & $0.93$ \\
C2 & $5 \times 10^{-5}$ & $4.6628 \times 10^6$ & $0.98$ & $2.5$ & $1$ & $128^3$ & $4.92$ & $52.5$ & $91\pm18$ & $0.041$ & $0.21$ & $0.30\pm0.10$ & $0.080$ & $1.80$ \\
C3 & $5 \times 10^{-5}$ & $4.5201 \times 10^6$ & $0.95$ & $2.5$ & $1$ & $128^3$ & $4.10$ & $41.4$ & $84\pm19$ & $0.042$ & $0.21$ & $0.27\pm0.10$ & $0.077$ & $1.95$\\
C4 & $5 \times 10^{-5}$ & $4.4250 \times 10^6$ & $0.93$ & $2.5$ & $1$ & $128^3$ & $4.92$ & $47.6$ & $81\pm17$ & $0.043$ & $0.21$ & $0.27\pm0.09$ & $0.078$ & $2.09$ \\
C5 & $5 \times 10^{-5}$ & $4.3298 \times 10^6$ & $0.91$ & $2.5$ & $1$ & $128^3$ & $3.28$ & $-$ & $-$ & $-$ & $-$ & $-$ & $-$ & $-$ \\
\hline
D1* & $2.5 \times 10^{-5}$ & $1.4081 \times 10^7$ & $1.18$ & $2.5$ & $1$ & $128^3$ & $1.64$ & $19.7$ & $170\pm35$ & $0.030$ & $0.16$ & $0.39\pm0.11$ & $0.068$ & $1.39$\\
D2 & $2.5 \times 10^{-5}$ & $1.1715 \times 10^7$ & $0.98$ & $2.5$ & $1$ & $128^3$ & $3.28$ & $31.9$ & $110\pm23$ & $0.036$ & $0.17$ & $0.30\pm0.09$ & $0.074$ & $2.51$ \\
D3 & $2.5 \times 10^{-5}$ & $1.1357 \times 10^7$ & $0.95$ & $2.5$ & $1$ & $128^3$ & $4.10$ & $37.2$ & $101\pm23$ & $0.037$ & $0.17$ & $0.27\pm0.10$ & $0.071$ & $2.71$ \\
D4 & $2.5 \times 10^{-5}$ & $1.1118 \times 10^7$ & $0.93$ & $2.5$ & $1$ & $128^3$ & $3.28$ & $28.9$ & $96\pm23$ & $0.037$ & $0.18$ & $0.27\pm0.09$ & $0.072$ & $2.95$\\
D5 & $2.5 \times 10^{-5}$ & $1.0640 \times 10^7$ & $0.91$ & $2.5$ & $1$ & $128^3$ & $2.73$ & $25.0$ & $86\pm21$ & $0.039$ & $0.18$ & $0.23\pm0.09$ & $0.070$ & $3.25$\\
D6 & $2.5 \times 10^{-5}$ & $1.0520 \times 10^7$ & $0.88$ & $2.5$ & $1$ & $128^3$ & $3.55$ & $28.7$ & $81\pm22$ & $0.039$ & $0.17$ & $0.21\pm0.10$ & $0.059$ & $3.23$\\
D7 & $2.5 \times 10^{-5}$ & $1.0161 \times 10^7$ & $0.85$ & $2.5$ & $1$ & $128^3$ & $1.04$ & $-$ & $-$ & $-$ & $-$ & $-$ & $-$ & $-$ \\
\hline
E1 & $1 \times 10^{-5}$ & $4.7712 \times 10^7$ & $1.18$ & $5$ & $1$ & $128^3$ & $0.98$ & $7.0$ & $340\pm87$ & $0.020$ & $0.12$ & $0.41\pm0.11$ & $0.042$ & $1.93$\\ 
E2 & $1 \times 10^{-5}$ & $3.9660 \times 10^7$ & $0.98$ & $5$ & $1$ & $128^3$ & $1.15$ & $4.9$ & $166\pm74$ & $0.027$ & $0.13$ & $0.25\pm0.14$ & $0.048$ & $5.46$ \\
E3 & $1 \times 10^{-5}$ & $3.8447 \times 10^7$ & $0.95$ & $5$ & $1$ & $128^3$ & $1.15$ & $4.8$ & $143\pm50$ & $0.028$ & $0.13$ & $0.22\pm0.11$ & $0.049 $ & $5.84$ \\
E4 & $1 \times 10^{-5}$ & $3.7672 \times 10^7$ & $0.93$ & $5$ & $1$ & $128^3$ & $1.15$ & $5.1$ & $145\pm44$ & $0.027$ & $0.13$ & $0.23\pm0.09$ & $0.049$ & $5.82$ \\
E5 & $1 \times 10^{-5}$ & $3.6810 \times 10^7$ & $0.91$ & $5$ & $1$ & $128^3$ & $7.22$ & $28.9$ & $129\pm46$ & $0.029$ & $0.13$ & $0.21\pm0.10$ & $0.045$ & $6.68$\\
E6 & $1 \times 10^{-5}$ & $3.6019 \times 10^7$ & $0.89$ & $5$ & $1$ & $128^3$ & $1.31$ & $-$ & $-$ & $-$ & $-$ & $-$ & $-$ & $-$\\
\hline
F1* & $1 \times 10^{-5}$ & $4.7712 \times 10^7$ & $1.18$ & $2.5$ & $1$ & $128^3$ & $1.80$ & $16.4$ & $196\pm42$ & $0.025$ & $0.12$ & $0.31\pm0.07$ & $0.057$ & $2.11$ \\
F2 & $1 \times 10^{-5}$ & $3.9660 \times 10^7$ & $0.98$ & $2.5$ & $1$ & $128^3$ & $1.05$ & $8.3$ & $129\pm29$ & $0.030$ & $0.13$ & $0.24\pm0.07$ & $0.060$ & $3.76$ \\
F3 & $1 \times 10^{-5}$ & $3.8447 \times 10^7$ & $0.95$ & $2.5$ & $1$ & $128^3$ & $2.59$ & $20.2$ & $125\pm30$ & $0.030$ & $0.13$ & $0.22\pm0.08$ & $0.055$ & $3.72$ \\
F4 & $1 \times 10^{-5}$ & $3.7672 \times 10^7$ & $0.93$ & $2.5$ & $1$ & $128^3$ & $2.07$ & $15.6$ & $116\pm26$ & $0.031$ & $0.13$ & $0.21\pm0.07$ & $0.055$ & $4.05$\\
F5 & $1 \times 10^{-5}$ & $3.6810 \times 10^7$ & $0.91$ & $2.5$ & $1$ & $128^3$ & $2.13$ & $15.7$ & $111\pm28$ & $0.031$ & $0.13$ & $0.20\pm0.07$ & $0.049$ & $4.16$ \\
F6 & $1 \times 10^{-5}$ & $3.6019 \times 10^7$ & $0.89$ & $2.5$ & $1$ & $128^3$ & $3.17$ & $21.6$ & $103\pm23$ & $0.031$ & $0.13$ & $0.19\pm0.06$ & $0.051$ & $4.44$\\
F7 & $1 \times 10^{-5}$ & $3.5209 \times 10^7$ & $0.87$ & $2.5$ & $1$ & $128^3$ & $3.28$ & $21.9$ & $97\pm26$ & $0.032$ & $0.13$ & $0.18\pm0.08$ & $0.044$ & $4.80$ \\
F8 & $1 \times 10^{-5}$ & $3.4400 \times 10^7$ & $0.85$ & $2.5$ & $1$ & $128^3$ & $0.82$ & $-$ & $-$ & $-$ & $-$ & $-$ & $-$ & $-$ \\
\hline
G1 & $1 \times 10^{-5}$ & $4.7712 \times 10^7$ & $1.18$ & $1$ & $1$ & $128^3$ & $1.64$ & $26.1$ & $108\pm16$ & $0.033$ & $0.13$ & $0.22\pm0.04$ & $0.068$ & $1.90$ \\
G2 & $1 \times 10^{-5}$ & $3.9660 \times 10^7$ & $0.98$ & $1$ & $1$ & $128^3$ & $1.31$ & $18.5$ & $81\pm13$ & $0.036$ & $0.14$ & $0.16\pm0.04$ & $0.059$ & $2.46$ \\
G3 & $1 \times 10^{-5}$ & $3.8447 \times 10^7$ & $0.95$ & $1$ & $1$ & $128^3$ & $1.48$ & $20.4$ & $76\pm12$ & $0.037$ & $0.14$ & $0.15\pm0.04$ & $0.058$ & $2.63$ \\
G4 & $1 \times 10^{-5}$ & $3.7672 \times 10^7$ & $0.93$ & $1$ & $1$ & $128^3$ & $1.31$ & $17.1$ & $70\pm15$ & $0.037$ & $0.13$ & $0.13\pm0.05$ & $0.053$ & $2.65$\\
G5 & $1 \times 10^{-5}$ & $3.6810 \times 10^7$ & $0.91$ & $1$ & $1$ & $128^3$ & $1.48$ & $19.6$ & $68\pm11$ & $0.037$ & $0.14$ & $0.12\pm0.03$ & $0.053$ & $2.66$ \\
G6 & $1 \times 10^{-5}$ & $3.6019 \times 10^7$ & $0.89$ & $1$ & $1$ & $128^3$ & $2.62$ & $33.6$ & $65\pm12$ & $0.038$ & $0.13$ & $0.11\pm0.04$ & $0.051$ & $2.70$\\
G7 & $1 \times 10^{-5}$ & $3.5209 \times 10^7$ & $0.87$ & $1$ & $1$ & $128^3$ & $0.16$ & $-$ & $-$ & $-$ & $-$ & $-$ & $-$ & $-$\\
\hline
H1* & $5 \times 10^{-6}$ & $1.2000 \times 10^8$ & $1.18$ & $2.5$ & $1$ & $128^3$ & $1.16$ & $8.8$ & $217\pm55$ & $0.022$ & $0.10$ & $0.25\pm0.07$ & $0.049$ & $2.89$ \\
H2 & $5 \times 10^{-6}$ & $1.0000 \times 10^8$ & $0.98$ & $2.5$ & $1$ & $128^3$ & $1.56$ & $10.7$ & $150\pm37$ & $0.026$ & $0.11$ & $0.20\pm0.06$ & $0.048$ & $4.70$ \\
H3 & $5 \times 10^{-6}$ & $9.5000 \times 10^7$ & $0.93$ & $2.5$ & $1$ & $128^3$ & $2.25$ & $17.2$ & $132\pm34$ & $0.027$ & $0.11$ & $0.18\pm0.07$ & $0.047$ & $5.43$ \\
H4 & $5 \times 10^{-6}$ & $9.3000 \times 10^7$ & $0.91$ & $2.5$ & $1$ & $128^3$ & $1.21$ & $8.2$ & $120\pm24$ & $0.028$ & $0.11$ & $0.17\pm0.05$ & $0.046$ & $5.93$ \\
H5 & $5 \times 10^{-6}$ & $9.0673 \times 10^7$ & $0.89$ & $2.5$ & $1$ & $128^3$ & $3.28$ & $20.3$ & $121\pm24$ & $0.027$ & $0.11$ & $0.16\pm0.05$ & $0.046$ & $5.70$ \\
H6 & $5 \times 10^{-6}$ & $8.8636 \times 10^7$ & $0.87$ & $2.5$ & $1$ & $128^3$ & $3.36$ & $-$ & $-$ & $-$ & $-$ & $-$ & $-$ & $-$ \\
\hline 
I1* & $5 \times 10^{-6}$ & $1.2000 \times 10^8$ & $1.18$ & $1$ & $1$ & $128^3$ & $1.72$ & $25.9$ & $131\pm20$ & $0.028$ & $0.11$ & $0.20\pm0.04$ & $0.057$ & $2.41$ \\
I2* & $5 \times 10^{-6}$ & $1.0000 \times 10^8$ & $0.98$ & $1$ & $1$ & $128^3$ & $1.64$ & $21.1$ & $99\pm16$ & $0.031$ & $0.11$ & $0.15\pm0.04$ & $0.051$ & $3.08$ \\
I3 & $5 \times 10^{-6}$ & $9.5000 \times 10^7$ & $0.93$ & $1$ & $1$ & $128^3$ & $2.18$ & $27.9$ & $90\pm18$ & $0.031$ & $0.11$ & $0.13\pm0.05$ & $0.050$ & $3.25$ \\
I4 & $5 \times 10^{-6}$ & $9.3000 \times 10^7$ & $0.91$ & $1$ & $1$ & $128^3$ & $1.64$ & $20.2$ & $88\pm18$ & $0.032$ & $0.11$ & $0.13\pm0.05$ & $0.048$ & $3.34$ \\
I5* & $5 \times 10^{-6}$ & $9.0673 \times 10^7$ & $0.89$ & $1$ & $1$ & $128^3$ & $2.28$ & $-$ & $-$ & $-$ & $-$ & $-$ & $-$ & $-$\\
\hline
J1 & $5 \times 10^{-7}$ & $2.5866 \times 10^9$ & $1.18$ & $1$ & $0.25$ & $64^2 \times 256$ & $0.14$ & $0.85$ & $149\pm97$ & $0.024$ & $0.058$ & $0.11\pm0.12$ & $0.039$ & $9.71$ \\
J2 & $5 \times 10^{-7}$ & $2.0824 \times 10^9$ & $0.98$ & $1$ & $0.25$ & $64^2 \times 256$ & $0.12$ & $0.74$ & $211\pm189$ & $0.027$ & $0.067$ & $0.29\pm0.39$ & $0.084$ & $24.63$ \\
J3 & $5 \times 10^{-7}$ & $2.0385 \times 10^9$ & $0.95$ & $1$ & $0.25$ & $64^2 \times 256$ & $0.11$ & $1.62$ & $218\pm155$ & $0.024$ & $0.089$ & $0.28\pm0.29$ & $0.033$ & $15.66$ \\
J4 & $5 \times 10^{-7}$ & $1.9947 \times 10^9$ & $0.91$ & $1$ & $0.25$ & $64^2 \times 256$ & $0.02$ & $-$ & $-$ & $-$ & $-$ & $-$ & $-$ & $-$ \\
\hline
\end{tabular}
\end{center}
\caption{Summary of the input and output quantities in the numerical simulations. $N_x$, $N_y$ and $N_z$ are the numerical resolution in $x$, $y$ and $z$.
$t_r$ is the total run time in units of a thermal diffusion timescale, $t_{\kappa}=d^2/\kappa$. $t_e$ is a characteristic timescale for the decay of the large-scale field due to turbulent diffusion from \citet{Cat91b}. In all cases the Prandtl number is kept constant at $\Pra=1$.
The asterisk next to the case number denotes the simulations performed in \citet{Stel04} for comparison. The output quantities (defined in \S\ref{sec:def}) are time averaged during the saturated phase.}
\label{table1}
\end{table*}

\subsection{Supercritical dynamos at moderate Ekman numbers}
\label{sec:super}

We first examine dynamo action in the supercritical range, that is for $\Ra>\Ra_c$.  The main characteristics of the flow and magnetic field in this system were previously described in SH04. Here we briefly recall some of these characteristics and identify the important ingredients for the subcritical behaviour (at $\Ra<\Ra_c$) that we will explore in the next section. Additionally, we present the generation mechanism of the coherent horizontally-averaged field, which we call the mean field. 
In this section, we focus primarily on a representative simulation at $\Ek=5\times 10^{-6}$, $\Pm=1$, $\Ra = 1.2 \times 10^8 = 1.18 \Ra_c$ and $\lambda=1$ (Case I1 in Table~\ref{table1}), which is typical of these supercritical dynamo calculations.

\subsubsection{Modification of the convection by the magnetic field}

\begin{figure}
\centering
\includegraphics[width=0.7\textwidth]{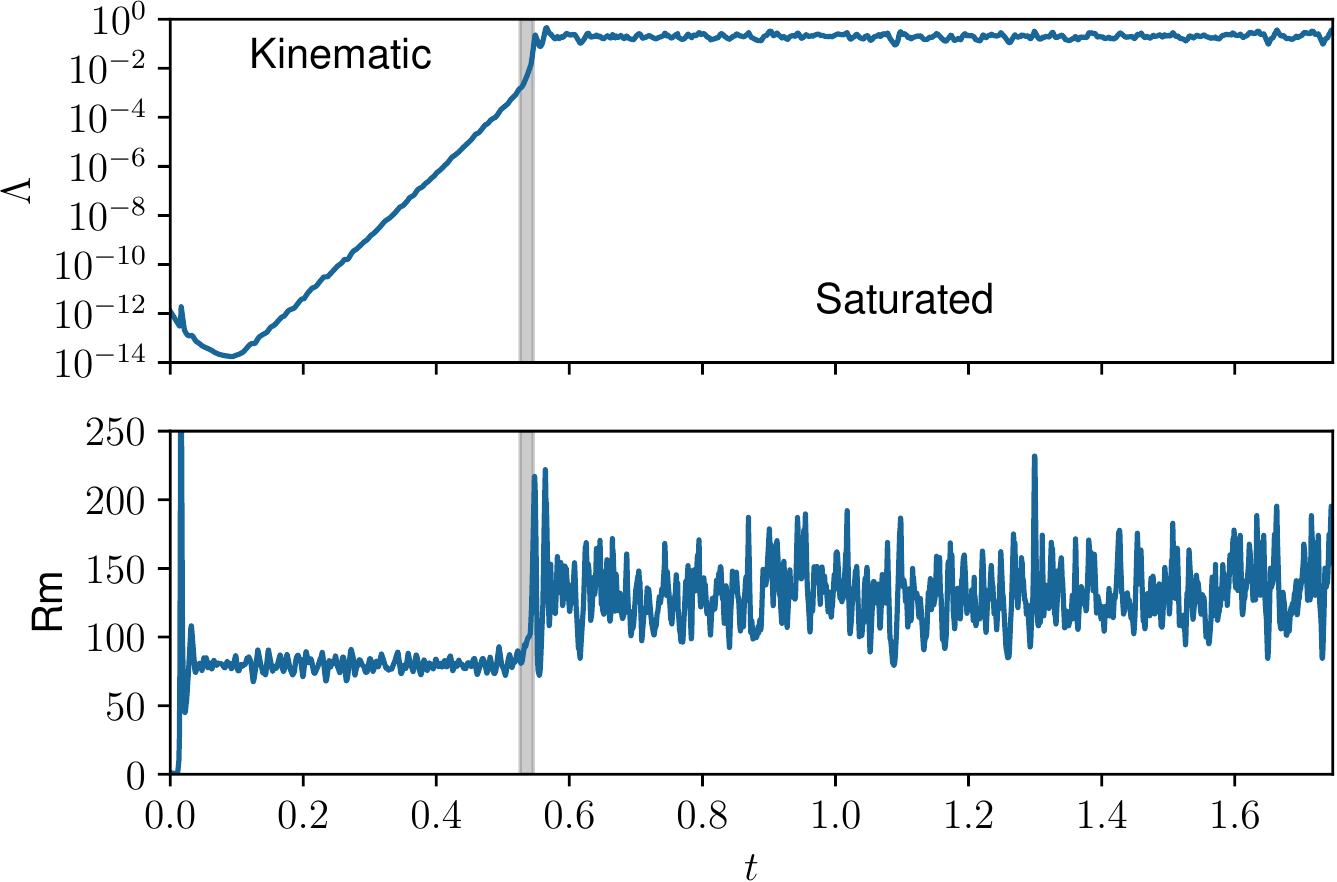}
\caption{Time series of the Elsasser number $\Lambda$ and magnetic Reynolds number $\Rm$ for Case I1. The grey area indicates the super-exponential growth phase.}
\label{fig:Run6_sup}
\end{figure}

For Case I1, the time evolution of the Elsasser number $\Lambda$ and magnetic Reynolds number $\Rm$ are shown in Figure \ref{fig:Run6_sup}. 
During the kinematic phase, where the magnetic energy grows exponentially and the Lorentz force does not significantly affect the flow, $\Rm$ takes an average value of $80$. Around $t=0.5$ in dimensionless units, the magnetic energy undergoes a short super-exponential growth phase. This phase occurs when the magnetic field reaches a sufficiently strong amplitude to alter the flow, leading to an amplification of the magnetic Reynolds number (and equivalently the kinetic energy), which is accompanied by an increasing growth rate of the magnetic energy due to the favourable effect on the dynamo action. Saturation then occurs and the magnetic field amplitude reaches an average value corresponding to $\Lambda\approx 0.2$. During the saturated phase, the magnetic Reynolds number exhibits greater fluctuations than in the kinematic phase and its average value is approximately $130$, which shows that the flow is significantly altered by the magnetic field. 
The ratio of magnetic to kinetic energies, $M$, is approximately $2-3$ in the saturated phase.
Values of $M>1$ and $\Lambda<1$ are typical of our saturated dynamos. For comparison, small-scale dynamos in rotating planar convection are typically characterised by a moderate reduction in $\Rm$ as the dynamo saturates and $M\sim O(0.1)$ \citep{Favier2011}. Some recent studies \citep[\eg][]{Sod12,Dor16} have adopted an alternative definition of the Elsasser number that is based on the typical velocity scale, $\Urms$, and the magnetic dissipation scale, $\lb$. In all cases, the modified Elsasser number, $\Lambda'=(\Ek/\Pra^2)\Brms^2/\Urms \lb$, is small, of the order of $10^{-2}$.

\begin{figure}
\centering
\subfigure[]{
\includegraphics[width=0.32\textwidth]{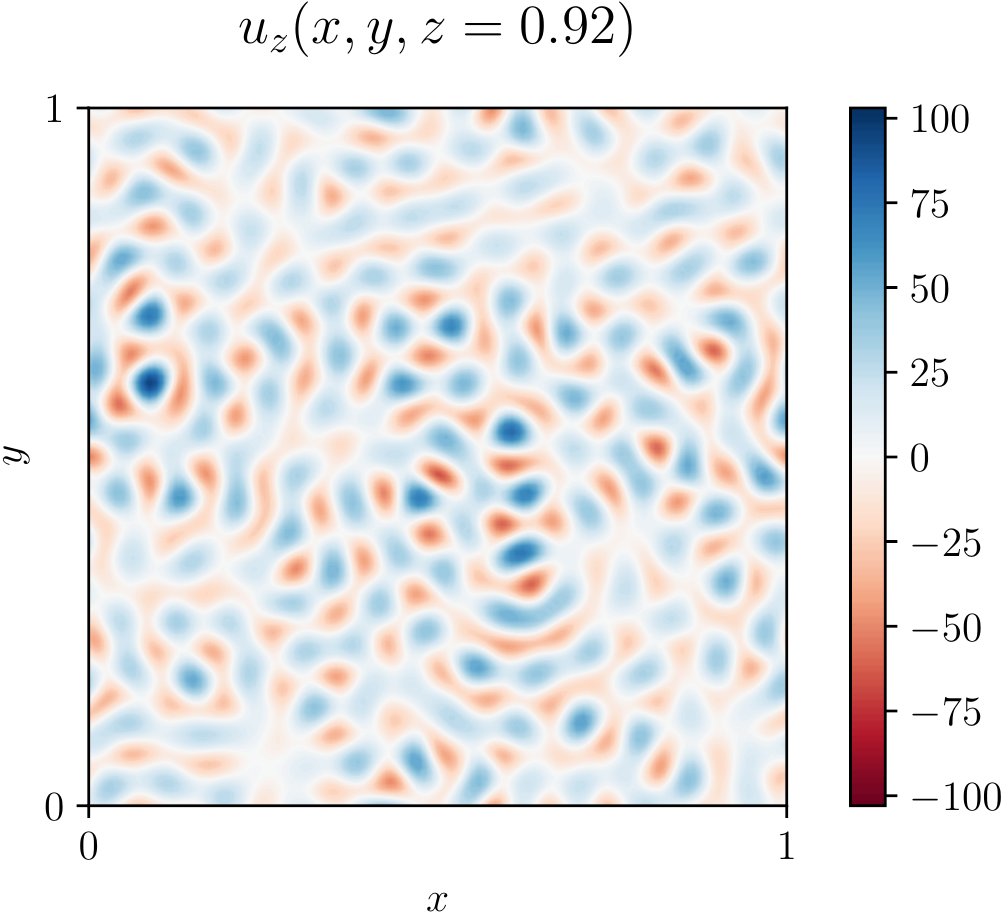}}
\subfigure[]{
\includegraphics[width=0.32\textwidth]{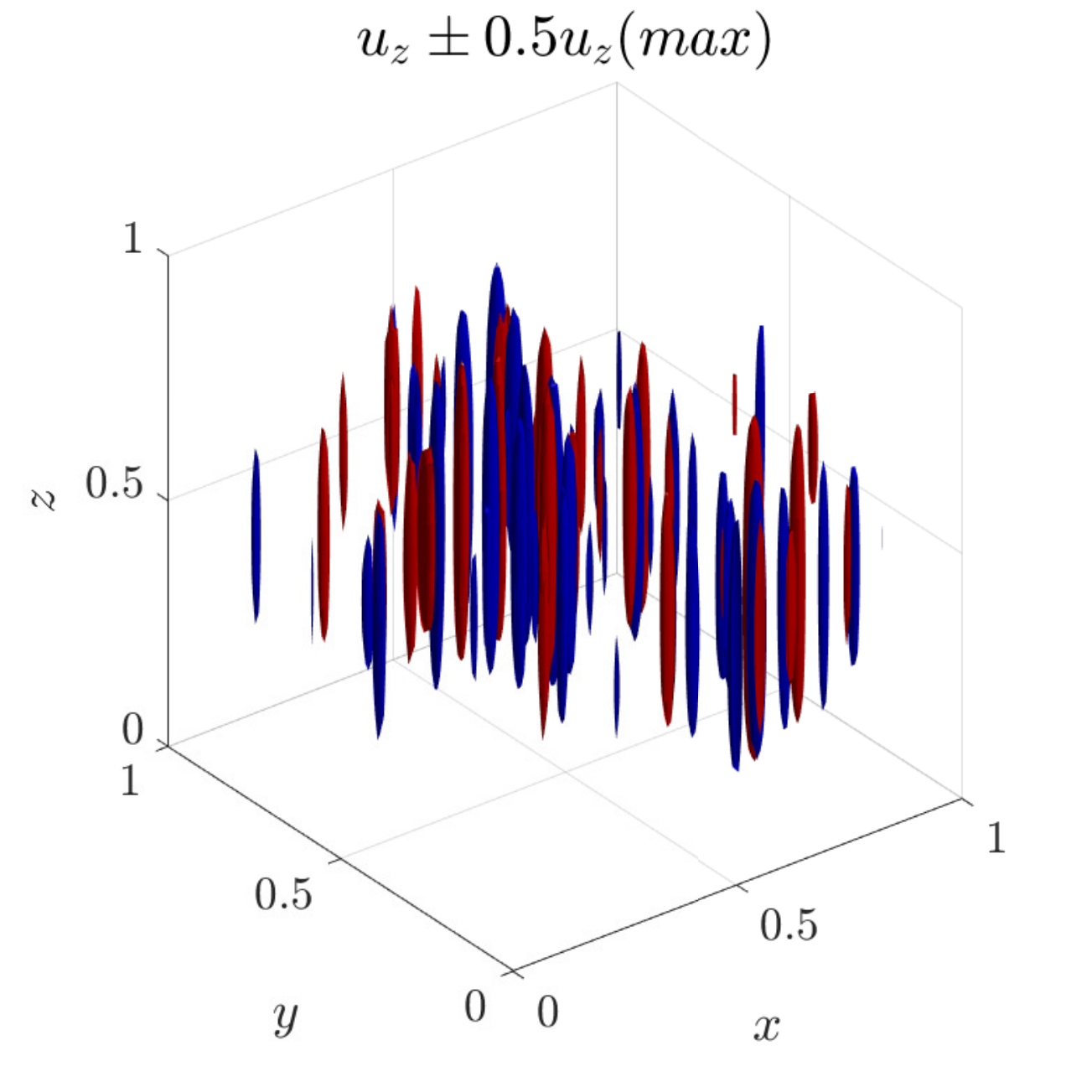}}
\subfigure[]{
\includegraphics[width=0.32\textwidth]{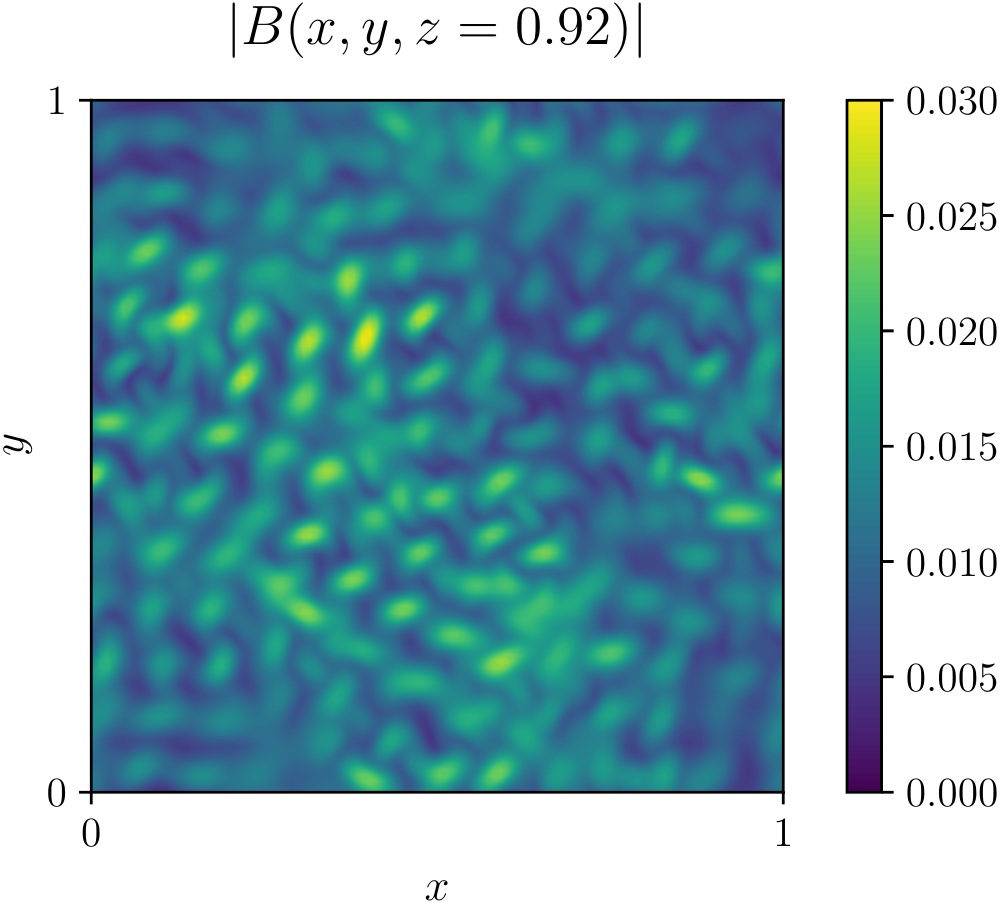}}
\subfigure[]{
\includegraphics[width=0.32\textwidth]{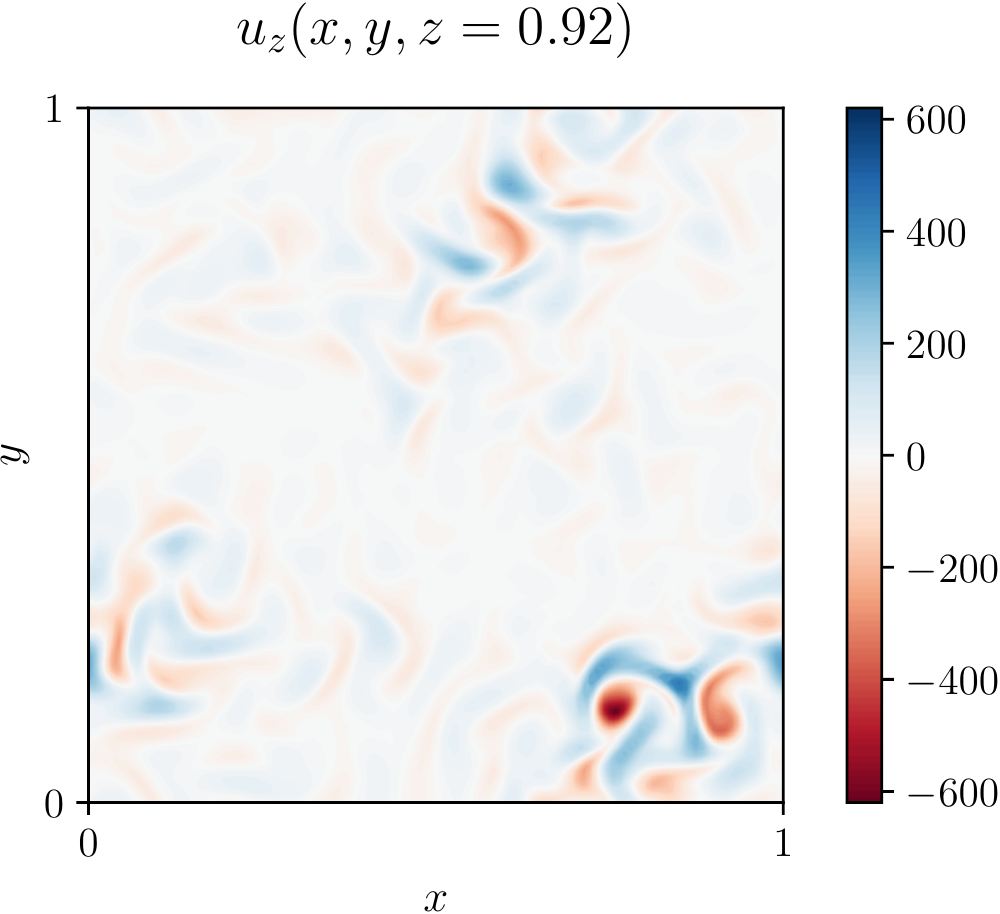}}
\subfigure[]{
\includegraphics[width=0.32\textwidth]{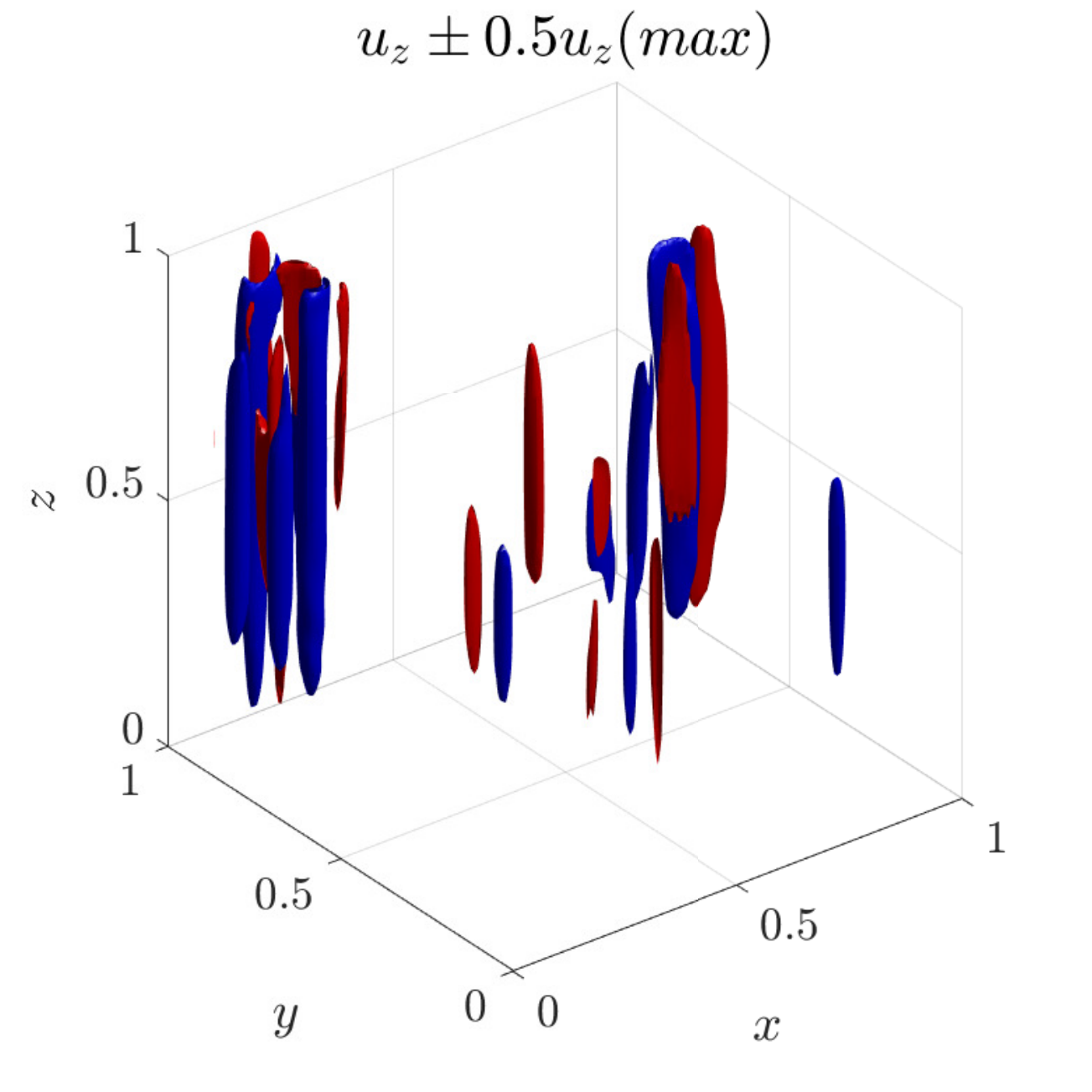}}
\subfigure[]{\label{fig:Bslice_sup}
\includegraphics[width=0.32\textwidth]{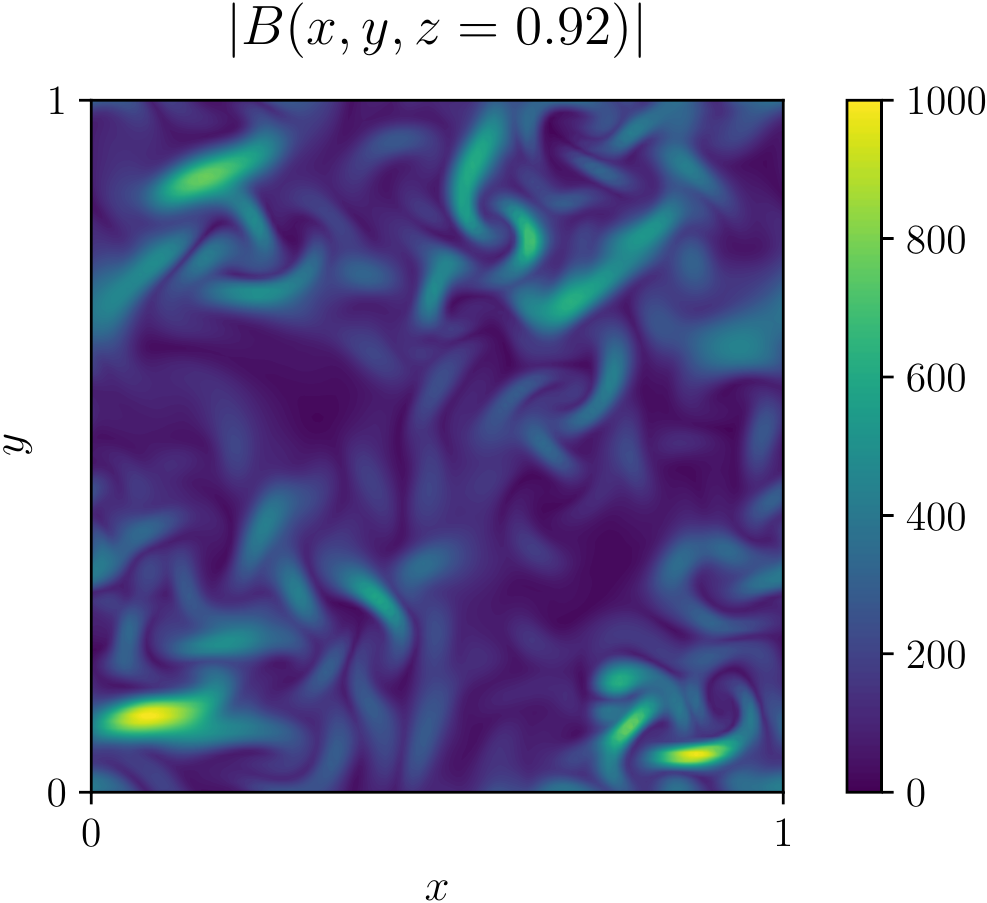}}
\caption{Snapshots in the (a-c) kinematic phase and (d-f) saturated phase of (a, d) $u_z$ in a horizontal slice, (b, e) 3D isosurfaces of $u_z$ and (c, f) the absolute value of the magnetic field, $|\bB|$, in a horizontal slice for Case I1. The horizontal slices are taken at $z=0.92$. The isosurfaces correspond to $\pm50\%$ of the maximum at that time.}
\label{fig:flow_sup}
\end{figure}

Changes in the flow between the kinematic and saturated phases can be observed in Figure~\ref{fig:flow_sup}, which shows horizontal slices and 3D isosurfaces of the vertical velocity. In the kinematic phase, the convection takes the form of tall and narrow columns that fill the space homogeneously. In the saturated phase, the convection becomes localised in a small number of large amplitude patches. The flow is still columnar but the columns appear fatter. This is confirmed by calculations of the horizontal flow lengthscale, which increases from a mean value of $0.08$ to $0.11$. In the kinematic phase, the vertical flow lengthscale is approximately box-sized (\ie close to unity) but decreases by approximately 20\% in the saturated phase, suggesting a weakening of the rotational constraints. This indicates that the convective columns become less depth-invariant as the dynamo saturates. The increase in the column width and the amplitude of the vertical velocity leads to an improved efficiency of the convective heat transport with an increase in the Nusselt number of 40\%. 

\begin{figure}
\centering
\subfigure[]{
\includegraphics[height=5.5cm]{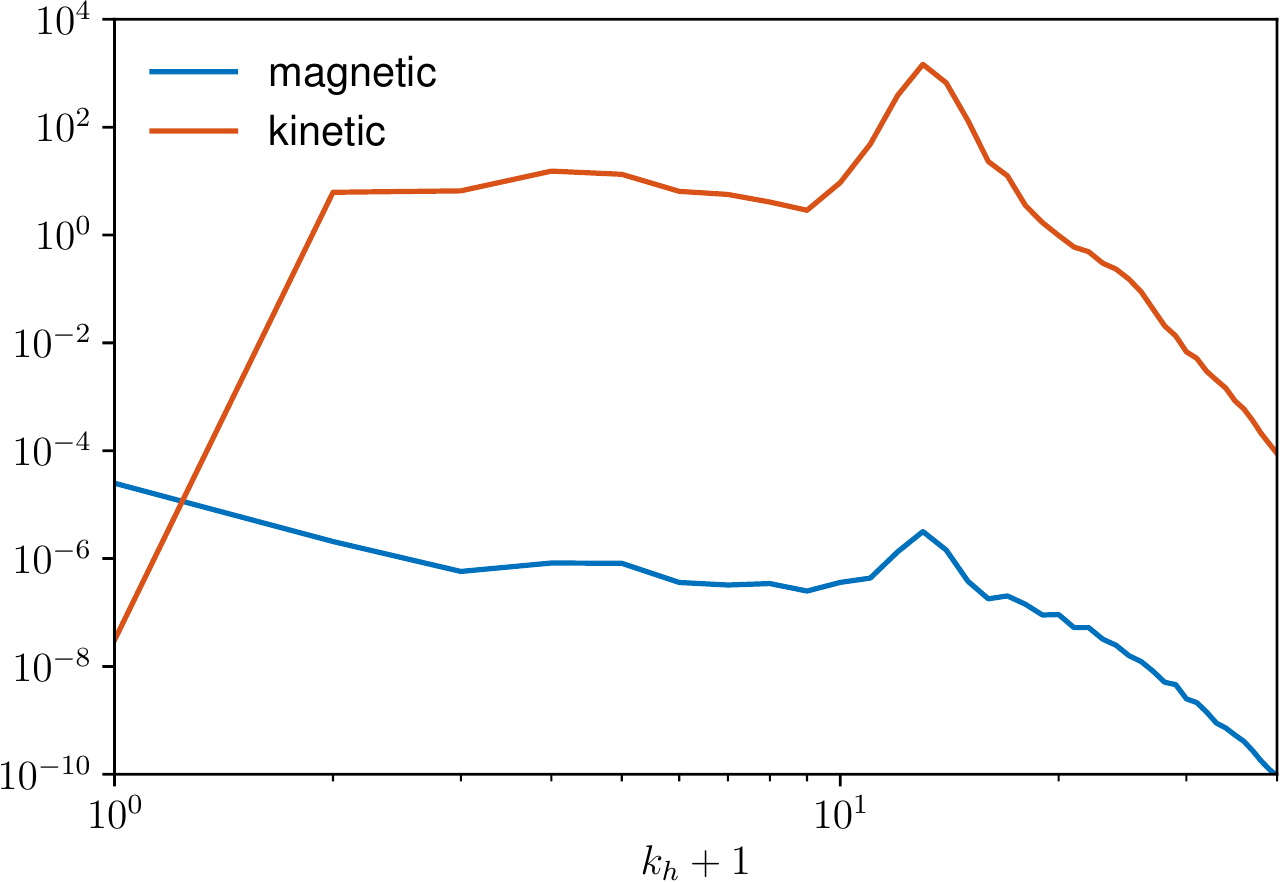}}
\subfigure[]{\label{fig:energy_spectra_saturated}
\includegraphics[height=5.5cm]{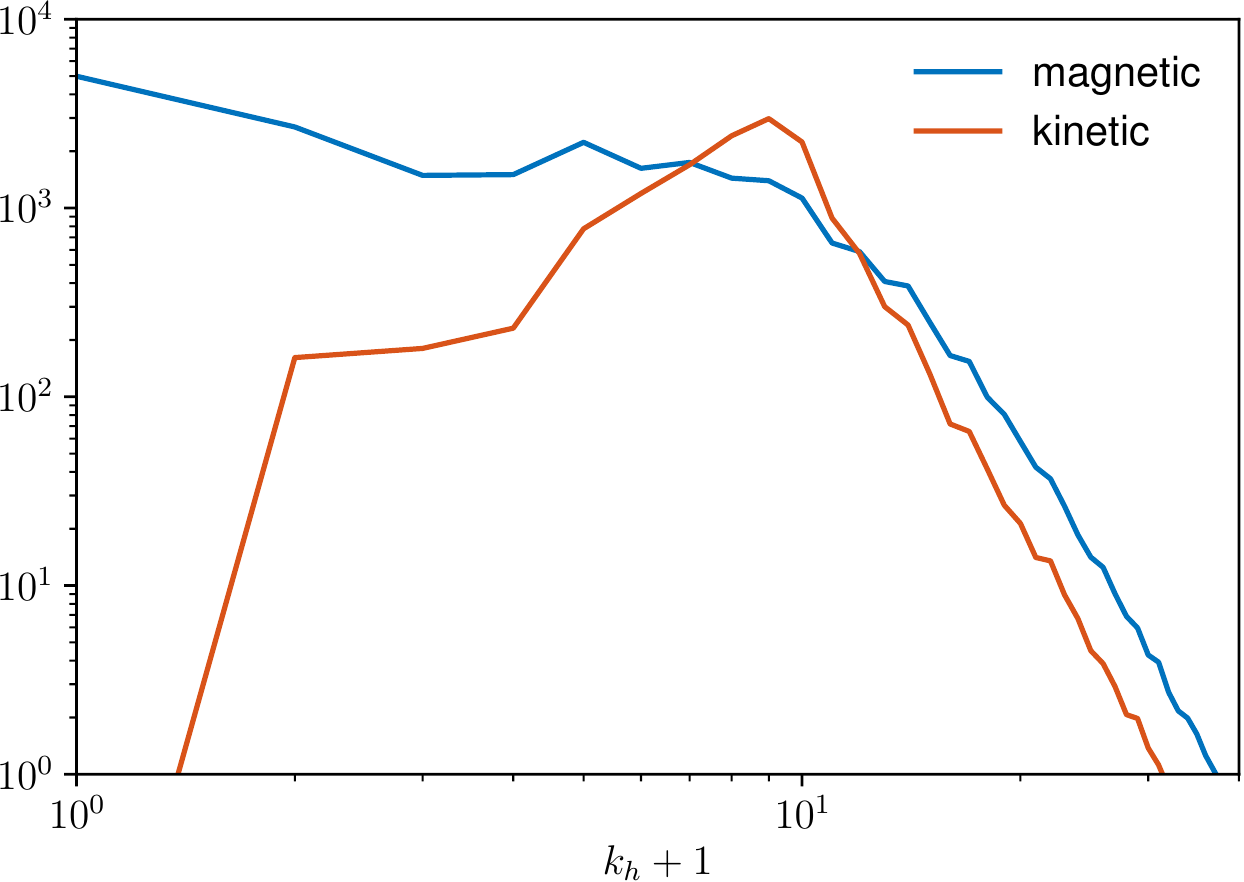}}
\caption{Power spectra (calculated from representative snapshots) of the magnetic energy and kinetic energy as a function of the horizontal wavenumber, $k_h=(k_x^2+k_y^2)^{1/2}$, during (a) the kinematic and (b) the saturated phases for Case I1. In order to include the contribution from the mean magnetic field (for which $k_h=0$) on the logarithmic scale, the horizontal wavenumbers are shifted to $k_h+1$.}
\label{fig:energy_spectra}
\end{figure}

\begin{figure}
\centering
	\subfigure[]{\label{fig:I1_meanfield_ts}
	\includegraphics[height=5cm]{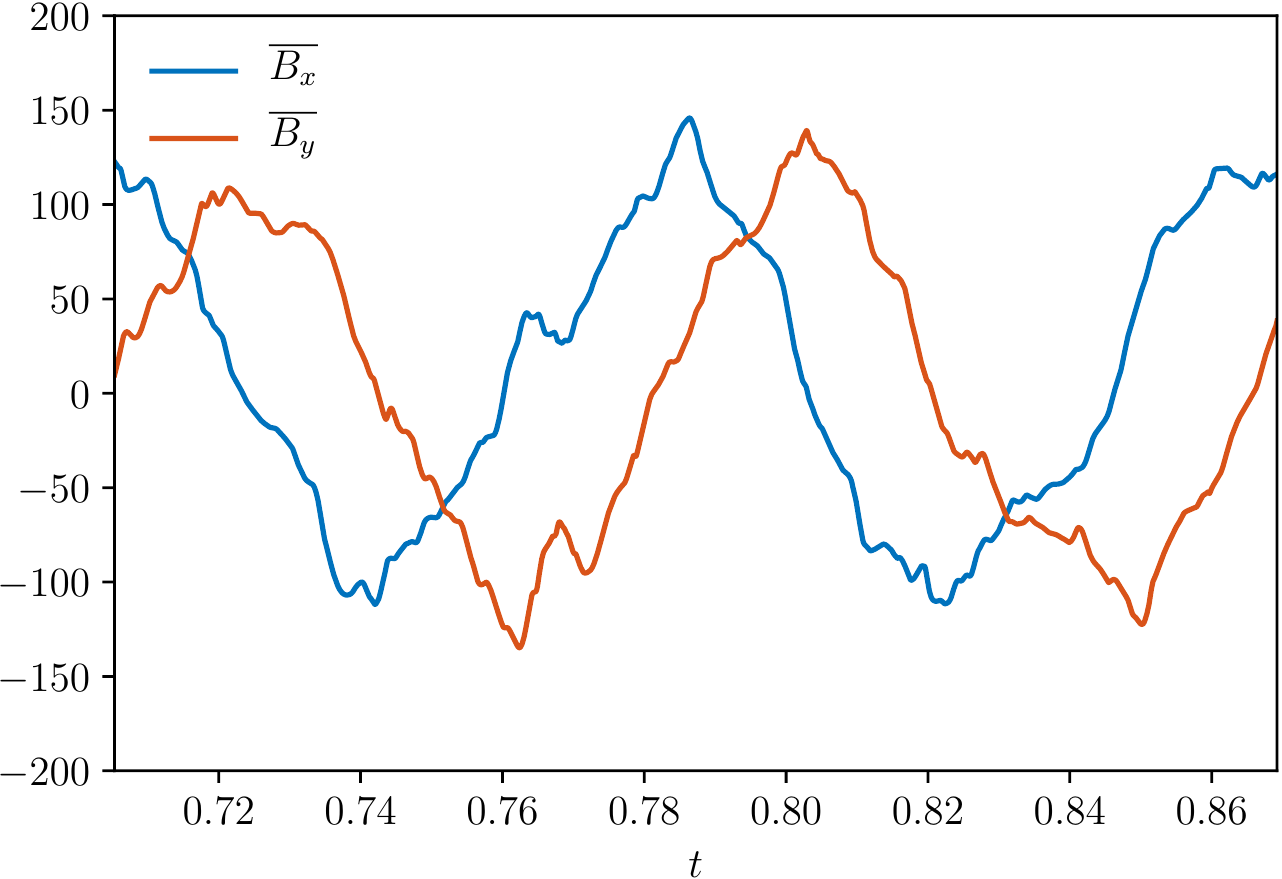}}
\subfigure[]{\raisebox{2.8mm}{\label{fig:I1_meanfield}
\includegraphics[height=4.75cm]{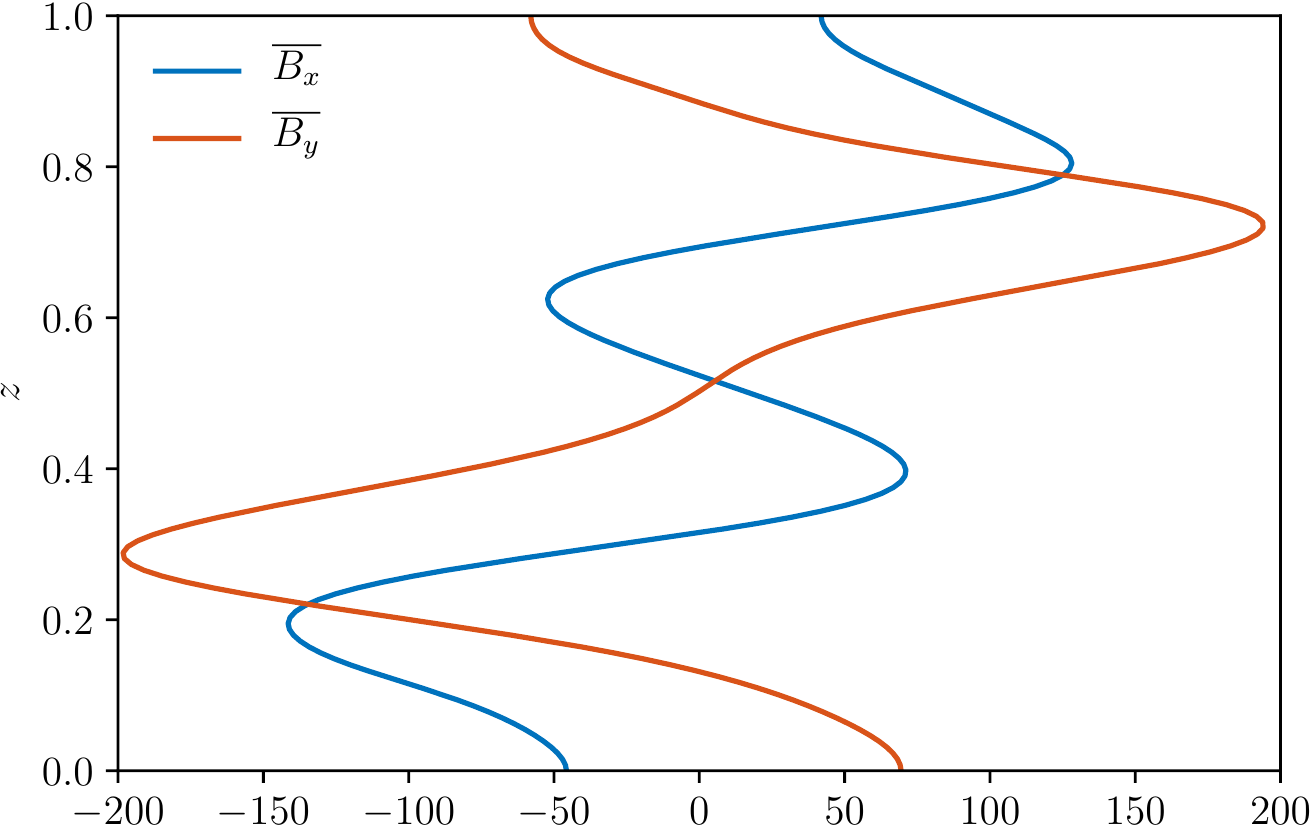}}}
\caption{(a) Time series of the $x$ and $y$ components of the mean magnetic field at $z=0.85$ and (b) snapshots of their vertical profile during the saturated phase for Case I1.}
\label{fig:B_meanfield}
\end{figure}

In order to observe the change in lengthscales as the dynamo saturates, Figure~\ref{fig:energy_spectra} shows the power spectra of the kinetic and magnetic energies in the two phases as a function of the horizontal wavenumber, $k_h=(k_x^2+k_y^2)^{1/2}$. The spectra are calculated at each depth and then averaged in $z$.
They are calculated from data snapshots but are a good representation of each phase.
The peak in the kinetic energy spectrum shifts from a horizontal wavenumber $k_h\approx12$ to $k_h\approx 8$, consistent with the change in the flow lengthscale. The peak is less pronounced in the saturated phase as a broad range of wavenumbers have significant energy. The magnetic field in the kinematic phase is dominated by the mean field (corresponding to $k_h=0$).  A secondary peak is also visible around the convection scale. In the saturated phase, the mean field is still dominant and contributes to approximately a quarter of the magnetic energy. The magnetic spectrum of the saturated phase is fairly flat from $k_h=1$ down to the convective scale. Figure~\ref{fig:Bslice_sup} shows that the fluctuating magnetic field (\ie the field corresponding to $k_h\geq 1$) is mainly concentrated in the regions of localised convection.  

\subsubsection{The mean magnetic field}

Focusing again upon case I1, Figure~\ref{fig:B_meanfield} shows the temporal evolution of the two horizontal components of the mean magnetic field alongside a snapshot of their vertical profiles in the saturated phase. In both the kinematic and the saturated phases, the mean field oscillates in time, generating the spiral staircase structure seen in SH04. Throughout its temporal evolution, the mean field remains approximately antisymmetric about the mid-plane ($z=0.5$) with a well-defined phase lag between the two horizontal components. The time period of oscillation is influenced by the parameters chosen. The magnetic Prandtl number has little effect on the time period of the mean field but higher $\Pm$ results in larger fluctuations, whereas at lower $\Ek$ the time period of oscillation decreases with increasing rotation rate.

In the plane layer dynamo model of \citet{Child72}, the dynamo operates as a two-scale mechanism, where the two scales involved are the box-size and the small horizontal scale that is associated with the near-onset convective flows. A small-scale magnetic field is produced by the distortion of a box-size (mean) field by the small-scale flow; the mean field is then produced by the mean electromotive force (e.m.f.) generated by the coherent interaction of small-scale field and flow. This dynamo process is thought to operate efficiently only close to the dynamo onset, where the Rayleigh number is small enough so that the convective motions are sufficiently laminar to cooperate in the production of the mean e.m.f. \citep{Courvoisier2009,Til12,Guer15}.
 
\begin{figure}
	\centering
	\subfigure[]{
		\includegraphics[width=0.45\textwidth]{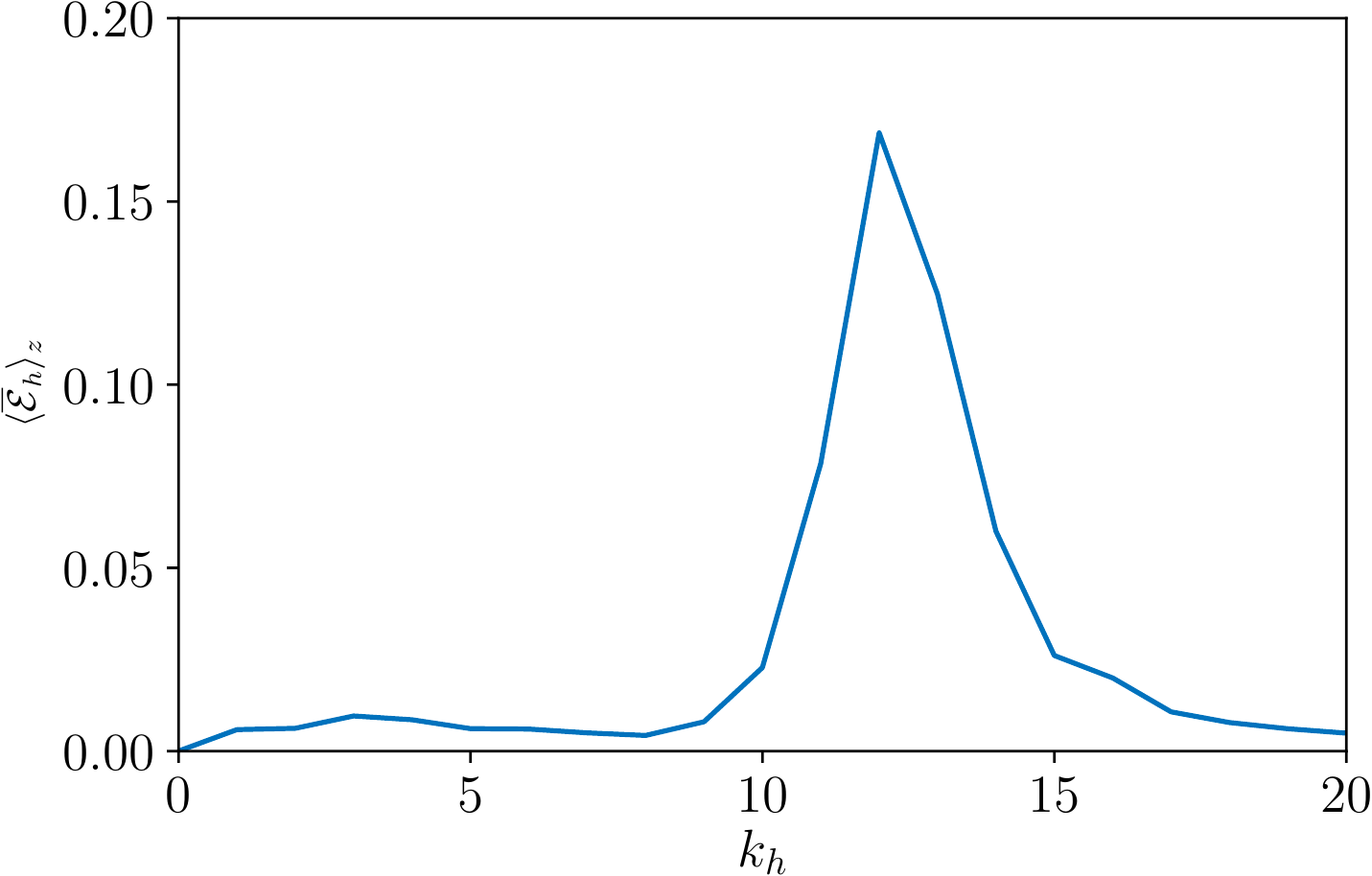}}
	\subfigure[]{
		\includegraphics[width=0.45\textwidth]{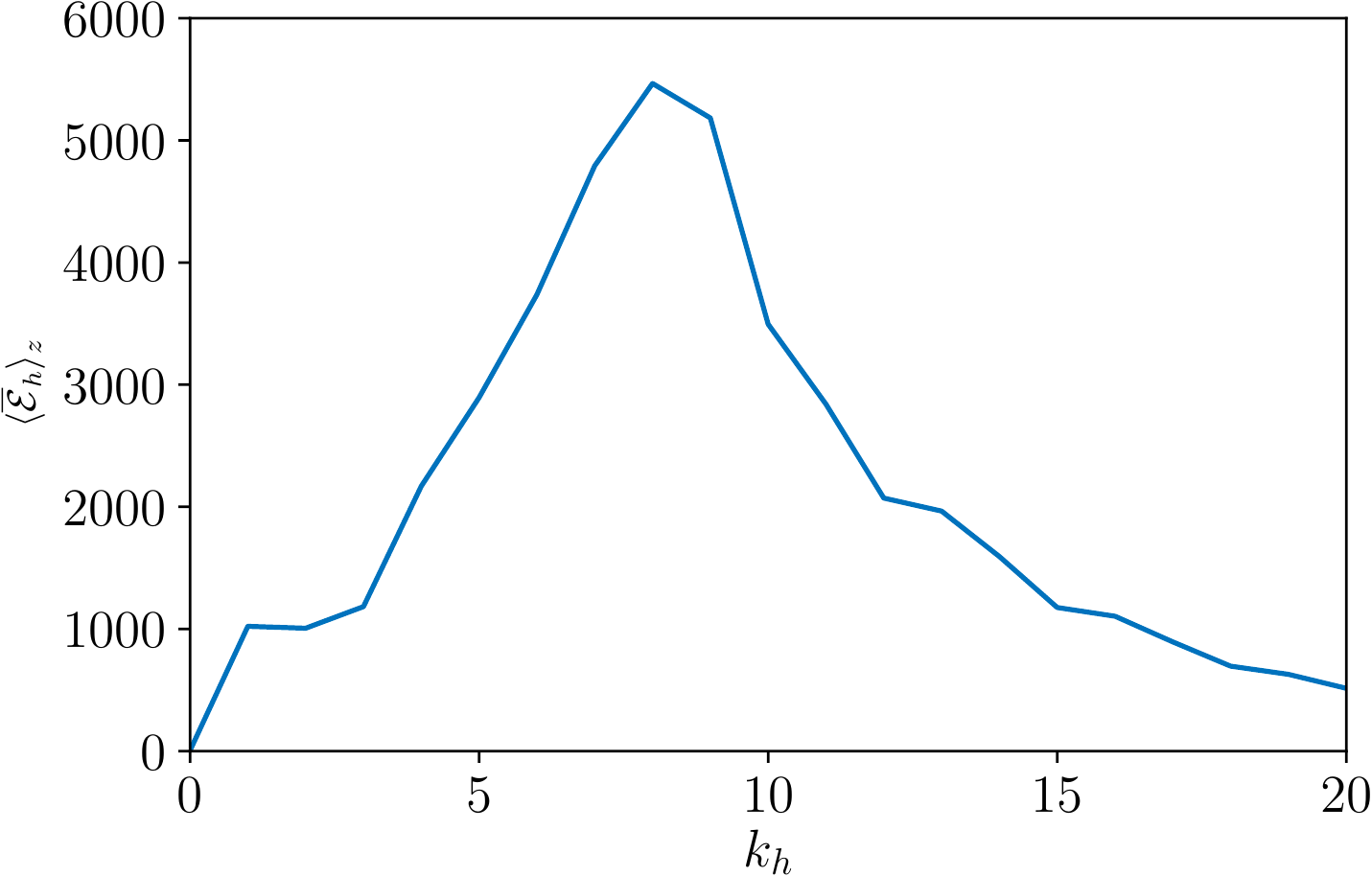}}
	\caption{Spectral contributions to the depth-averaged horizontal mean e.m.f., $\langle\overline{\mathcal{E}}_h\rangle_z=\langle(\overline{\mathcal{E}}_x^2+\overline{\mathcal{E}}_y^2)^{1/2}\rangle_z$, taken from representative snapshots in (a) the kinematic phase and (b) the saturated phase of Case I1. Here $\langle ... \rangle_z$ denotes averaging in $z$.} 
	\label{fig:emf_spectra}
\end{figure}

\begin{figure}
\centering
\subfigure[]{
\includegraphics[width=0.45\textwidth]{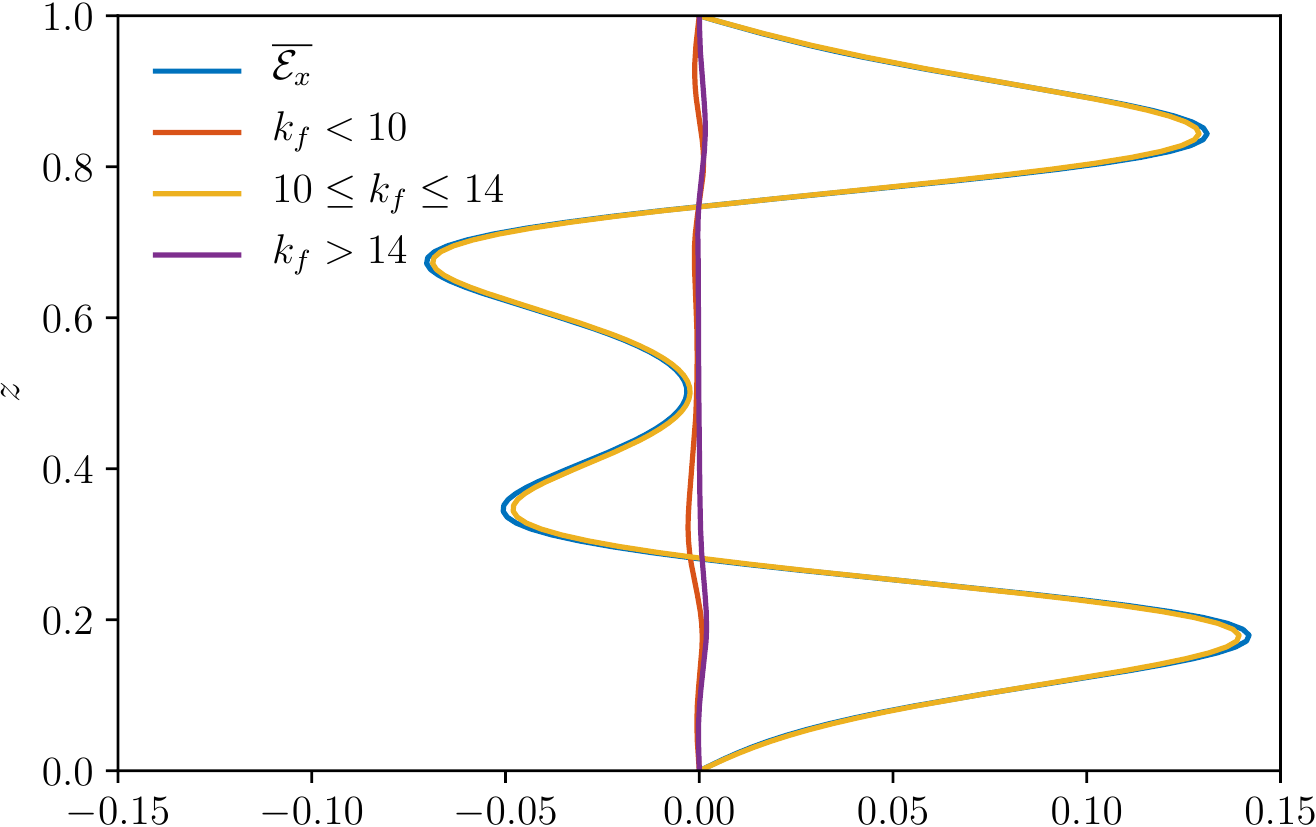}}
\subfigure[]{
\includegraphics[width=0.45\textwidth]{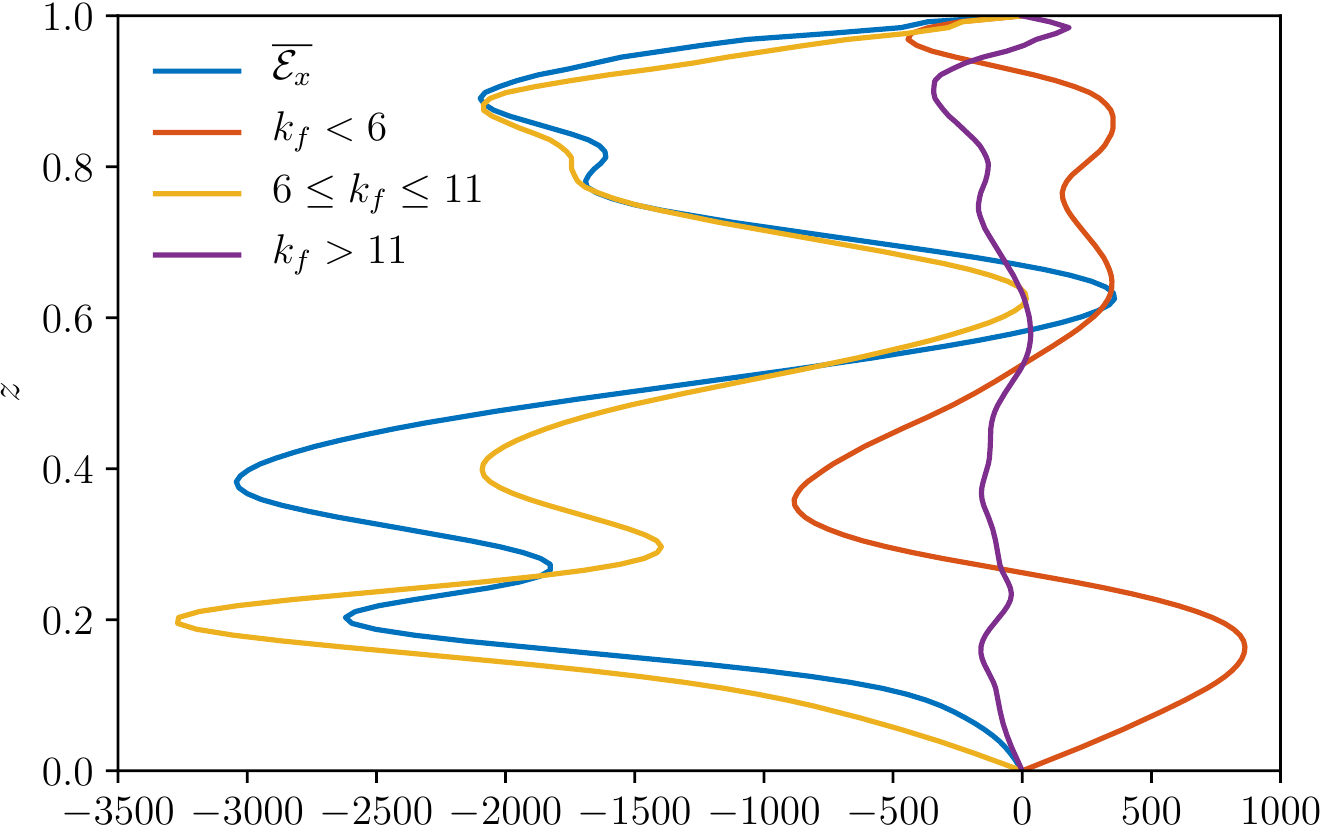}}
\caption{Vertical profile of the unfiltered and filtered versions of $\overline{\emf_x}$ in (a) the kinematic phase and (b) the saturated phase (snapshot) for Case I1 where contributions from $k_h=k_f$ are kept and other wavenumbers are filtered out.} 
\label{fig:emf}
\end{figure}

We want to establish whether the dynamo in these simulations operates along the same principle as the two-scale dynamo in the saturated phase. We anticipate that the picture might be more complicated in our simulations because: (i) although the kinetic energy spectra has a fairly well defined convective scale $k_h\approx8$ (Figure~\ref{fig:energy_spectra_saturated}), the intermediate scales (\ie the scales between the box size and the convective scale) have significant energy; (ii) the magnetic energy spectrum tends to be flat at the intermediate scales, without a clear peak at the convective scale. Following \citet{Bush18}, we study the contributions to the mean e.m.f., which is responsible for the generation of the mean field, $\overline{\bemf} = \overline{\bu \times \bB}$.
To investigate the contributions of different scales of motion to this mean e.m.f., we can filter the velocity field in Fourier space into contributions from each mode. Note that any flow filtering is done post-processing from representative data snapshots, so the dynamo calculation is not affected by the filtering process.
We study the respective contributions from individual horizontal wavenumbers to the mean horizontal e.m.f., $\overline{\mathcal{E}}_h$. We calculate $\overline{\mathcal{E}}_h=(\overline{\mathcal{E}}_x^2+\overline{\mathcal{E}}_y^2)^{1/2}$ as a function of depth, from these filtered flows, before depth-averaging this quantity. This depth-averaged value is then plotted in Figure~\ref{fig:emf_spectra}. In the kinematic phase, we observe a dominant contribution to the mean e.m.f. produced by $k_h\sim12$ with little contribution from lower or higher wavenumbers, indicating a two-scale mechanism. These ``active'' wavenumbers (\ie those producing the largest spectral contributions) shift to lower values in the saturated phase and approximately corresponds to the dominant convective scale. 
A wider range of active wavenumbers exist in the saturated phase.
Figure~\ref{fig:emf} shows the unfiltered and filtered versions of the $x$-component of the mean e.m.f., $\overline{\emf_x}$, where we consider the depth-dependence of the mean e.m.f. produced by low, active and high wavenumbers. In the kinematic phase, the mean e.m.f. is almost entirely produced by contributions from the convective scales at $10 \leq k_h\leq 14$ with little contribution from lower or higher wavenumber. In the saturated phase, the mean e.m.f is mostly produced by motions at wavenumbers $6 \leq k_h \leq 11$, although the contribution from lower wavenumbers is more sizeable in this phase. These results suggest that the production of the coherent mean e.m.f. is dominated by motions at the convective scale, especially in the kinematic phase. Therefore the dynamo does indeed operate as a two-scale process.

\subsubsection{Variation in the dominant force balances}
\begin{figure}
	\centering
	\subfigure[]{
		\includegraphics[width=0.45\textwidth]{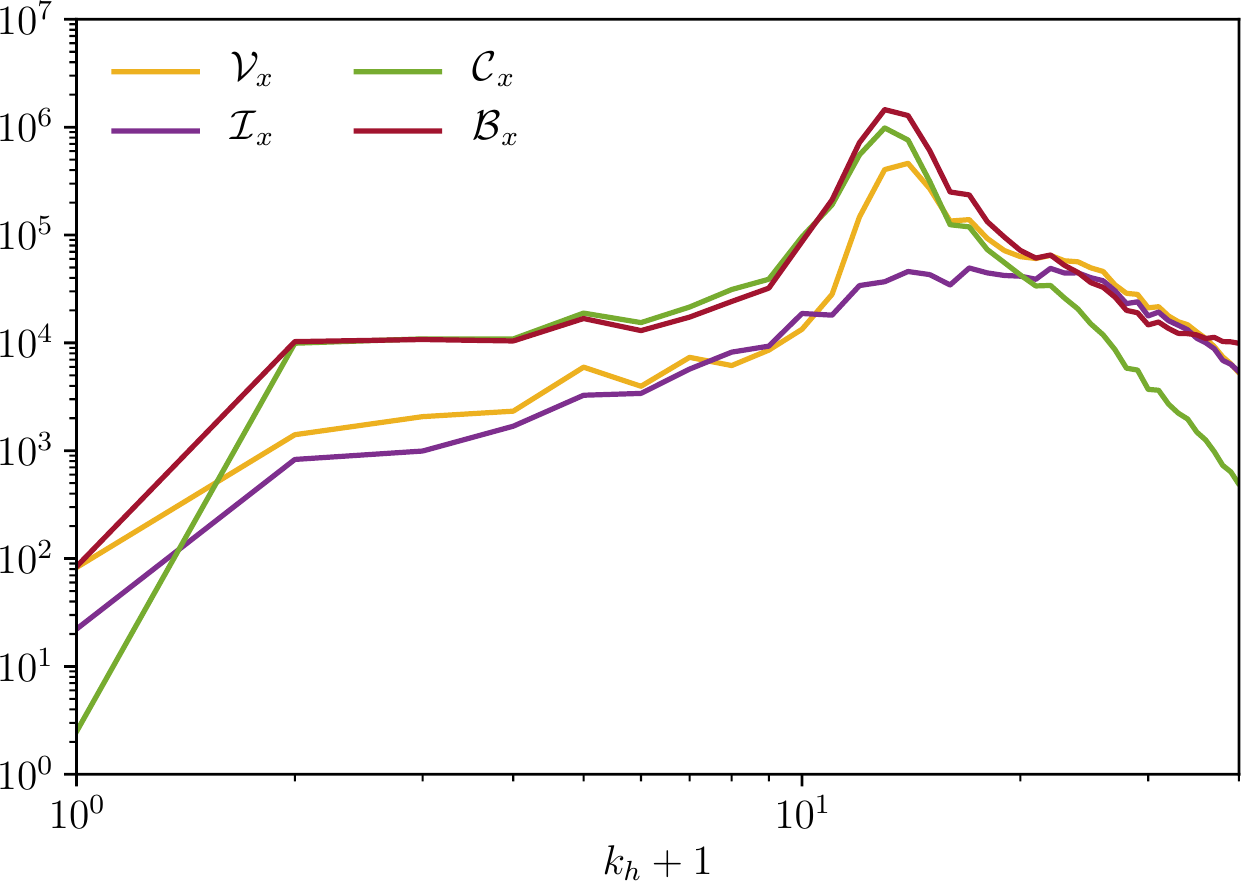}}
	\subfigure[]{
		\includegraphics[width=0.45\textwidth]{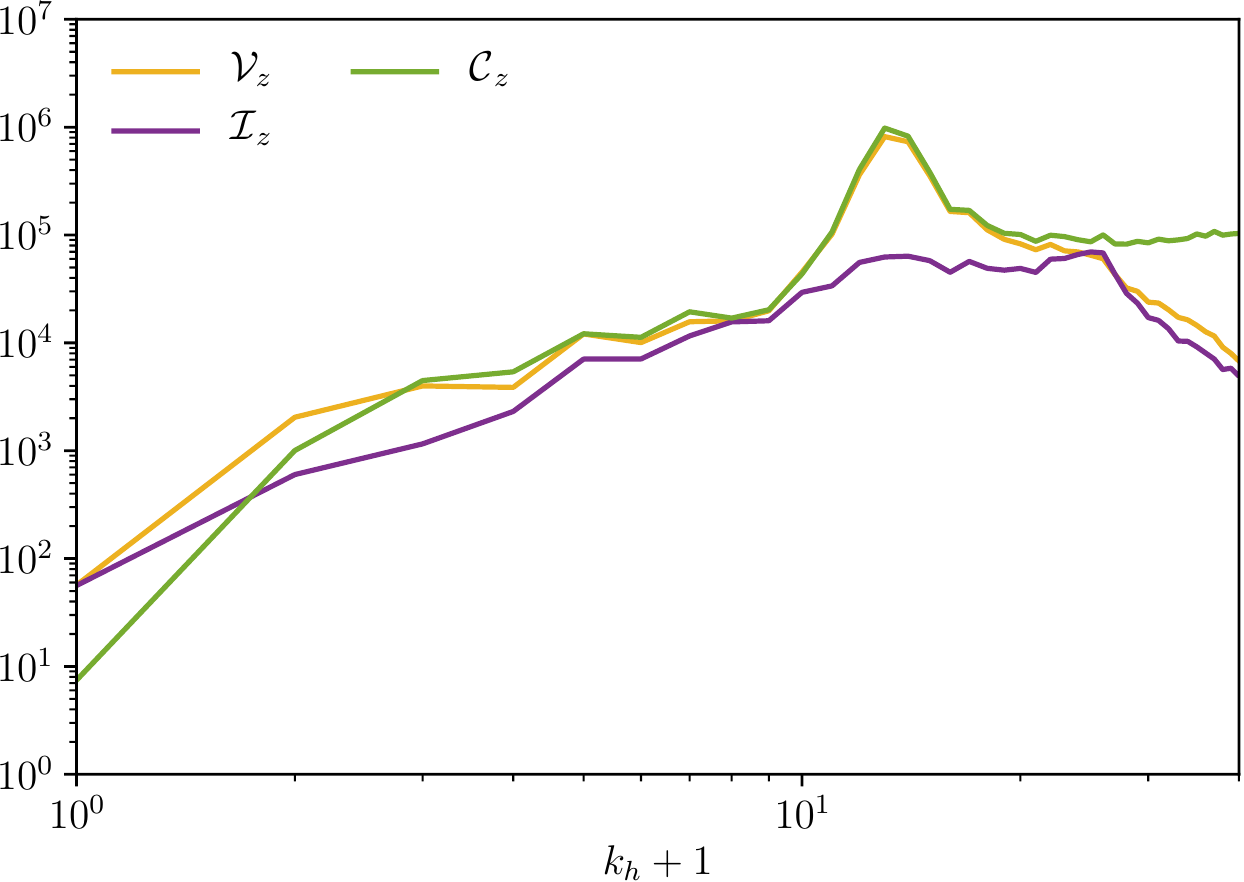}}\\
	\subfigure[]{
		\includegraphics[width=0.45\textwidth]{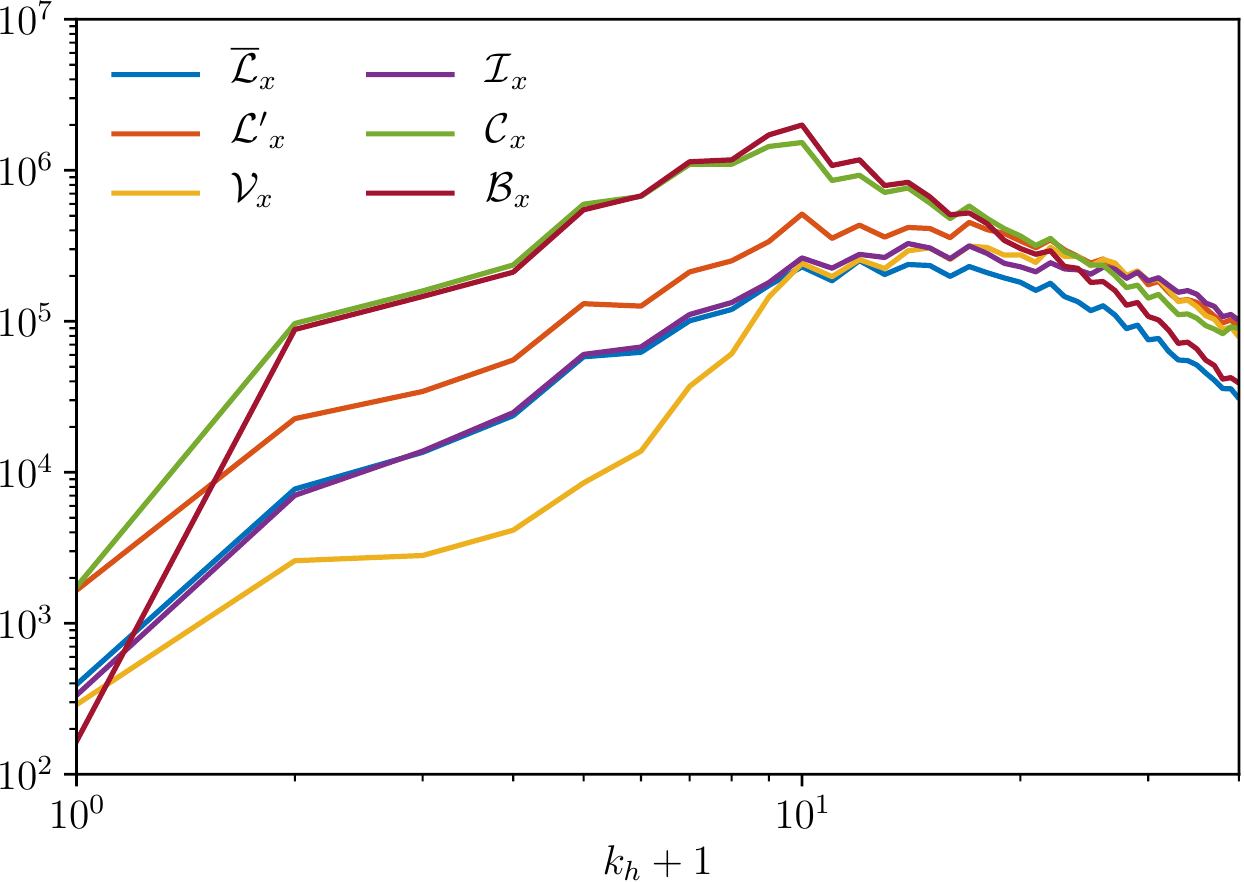}}
	\subfigure[]{
		\includegraphics[width=0.45\textwidth]{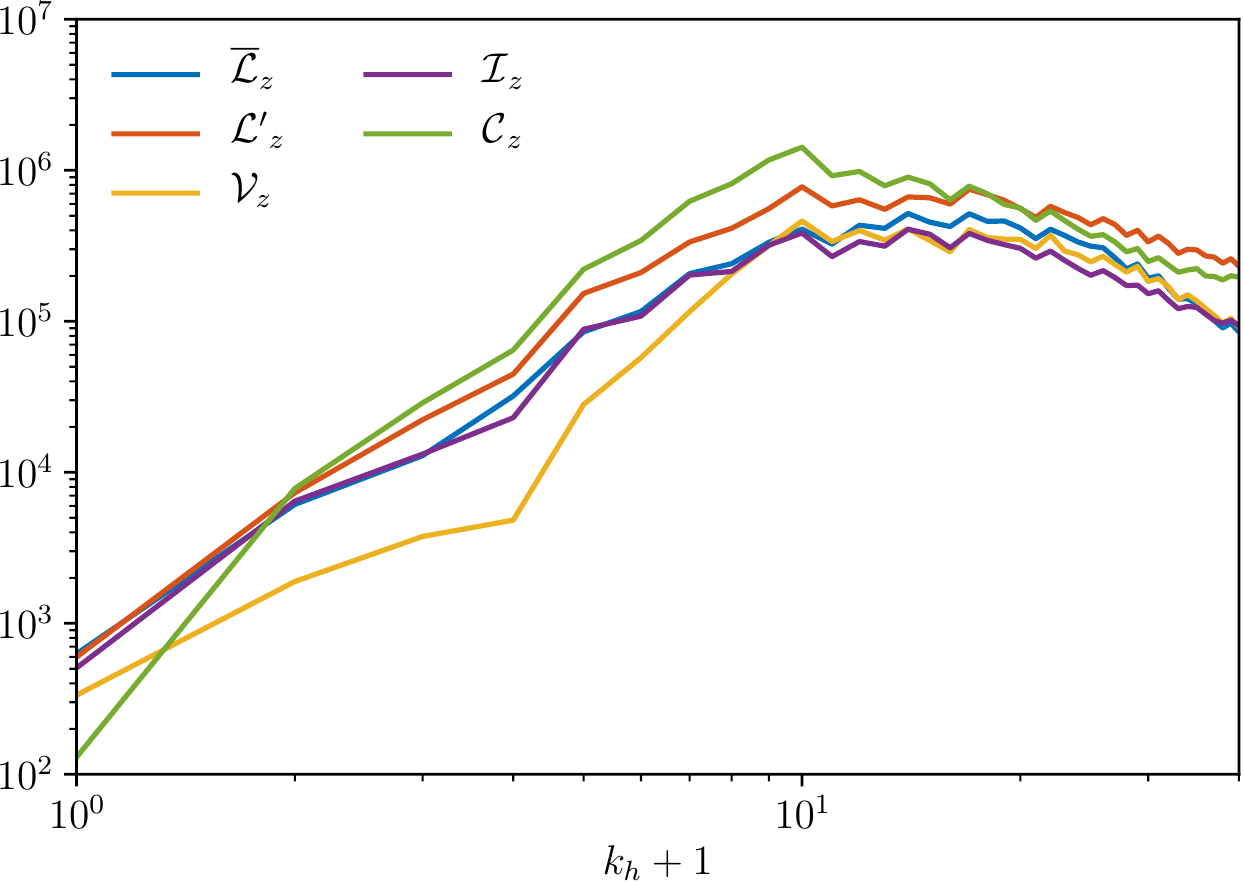}}
	\caption{Spectral distributions of the terms generating the (left) horizontal and (right) vertical vorticity as a function of the horizontal wavenumber for representative snapshots of Case I1 in the (a-b) kinematic and (c-d) saturated phases. The nonlinear inertial terms are denoted by $\mathcal{I}=(\bu \cdot \del) \vor - (\vor \cdot \del) \bu$, the Coriolis term by $\mathcal{C}=(\Pra/\Ek)\partial_z \bu$, the viscous term by $\mathcal{V}=\Pra \del^2 \vor $, the buoyancy term by $\mathcal{B}=\Pra \Ra \del \times (\Tpert \be_z)$, the Lorentz force term involving the mean field by $\overline{\mathcal{L}}$, and the Lorentz force term involving the fluctuating field only by $\mathcal{L}^\prime$.}
	\label{fig:force_spectra}
\end{figure}

The dominant force balances involved in dynamo action at low $\Ek$ are a source of wide speculation due to numerical restrictions on simulations \citep[e.g.][]{Cat17,Schw19}. Here we compare the strength of the different terms in the vorticity equation at different horizontal lengthscales in order to understand how the field modifies the flow and the relevant force balances, including any variations in these balances across different scales. Using the vorticity equation is a convenient way to eliminate the pressure terms which form a geostrophic balance with part of the Coriolis force. 
In dimensionless form, the vorticity equation is
\begin{equation}
\frac{\partial \vor}{\partial t} + (\bu \cdot \del) \vor - (\vor \cdot \del) \bu - \frac{\Pra}{\Ek} \frac{\partial \bu}{\partial z}  
= \Pra \del^2 \vor + \Pra \Ra \del \times (\Tpert \be_z) + \del \times (\bJ \times \bB).
\end{equation}
The spectral distribution of each term in either the $x$- or $z$-components of the vorticity equation, as a function of the horizontal wavenumber, is given in Figure~\ref{fig:force_spectra}. Here we show representative snapshots of the $x$-component of the vorticity, however analysis in the $y$-direction yields similar results. Each term is calculated from a snapshot of the velocity, magnetic field and temperature, which is then transformed into Fourier space at each depth, before being squared and then averaged along $z$.
In order to determine the effects of the mean and fluctuating parts of the magnetic field separately, the term corresponding to the curl of the Lorentz force is separated into the contributions from the fluctuating field only (i.e. the modes $k_h\geq1$, which we denote by a prime) $\mathcal{L}^\prime=\del \times \left(\bJ' \times \bB'\right)$, and the contributions involving the mean field $\overline{\mathcal{L}}=\del \times (\overline{\bJ}\times \overline{\bB}+\overline{\bj}\times\bB'+\bj' \times \overline{\bB})$. The remaining terms are the nonlinear inertial terms $\mathcal{I}=(\bu \cdot \del) \vor - (\vor \cdot \del) \bu$, the contribution from the Coriolis force $\mathcal{C}=(\Pra/\Ek)\partial_z \bu$, the viscous term $\mathcal{V}=\Pra \del^2 \vor $, and the term due to buoyancy $\mathcal{B}=\Pra \Ra \del \times (\Tpert \be_z)$.

In the kinematic phase, the main balance for the horizontal vorticity is between the buoyancy term and the Coriolis term at the convective scale, with a smaller contribution from the viscous term. A similar balance is also observed at larger scales. The inertial terms are significantly smaller, but not completely negligible even though $\Ra$ is only just above $\Ra_c$.  There is no buoyancy source for the vertical vorticity, so the Coriolis term is balanced by the viscous term at the convective scale. This picture is similar to the force balance expected at the linear onset of non-magnetic rotating convection \citep{Chandra61}. After comparing the force balances of multiple snapshots during the temporal evolution, this particular snapshot is a good representation of the dynamics during this phase.
In the saturated phase, the main balance for the horizontal vorticity is still between the buoyancy and Coriolis terms. In the vertical vorticity equation, the Coriolis term is now mostly balanced by the Lorentz term involving the interactions of the fluctuating field. This balance occurs at a larger wavenumber, where the viscous term is smaller. 
This picture differs somewhat from the force balances expected in linear rotating magnetoconvection \citep{Chandra61,Elt72}, where it is the linear Lorentz forces (i.e. the interaction between the mean and fluctuating fields) that balances the pressure gradient and the Coriolis force.
Nevertheless, the presence of a coherent mean field is crucial for the flow changes that we observe because the values of the magnetic Reynolds numbers obtained in these near-onset dynamos are not large enough to produce small-scale dynamo action \citep{Til12,Guervilly2017a}.  
The generation of the fluctuating field therefore relies on the presence of the mean field.
The mean field does not need to be particularly strong in terms of an Elsasser number: values of $\overline{\Lambda}\approx0.05$ are sufficient. The threshold value of $\overline{\Lambda}$ for which flow changes are significant will be explored in \S\ref{sec:sub}.

\subsection{Subcritical dynamos at moderate Ekman numbers}
\label{sec:sub}

Subcritical behaviour, where convection and dynamo action are sustained for $\Ra<\Ra_c$, is made possible in this system due to the significant changes in the flow that are induced by the presence of a non-negligible magnetic field.  Indeed, SH04 have already obtained a probable subcritical dynamo for a slightly subcritical Rayleigh number, $\Ra/\Ra_c=0.98$ for $\Ek=5\times10^{-6}$ and $\Pm=1$ (Case I2). However, as they pointed out, an infinitely long simulation would be needed to categorically prove the persistence of the subcritical behaviour in direct numerical simulations. For practical reasons, these simulations cannot be performed for more than a few magnetic diffusion timescales, but persistent magnetic field generation for the duration of a long calculation is strongly indicative of a subcritical dynamo. The running time $t_r$ is given in Table~\ref{table1}. As times are scaled with $d^2/\kappa$, $t_r \Pra/\Pm$ gives the running time in units of the ohmic decay time, $t_{\eta}=d^2/\eta$. The subcritical cases are run for somewhere between $t_{\eta}$ and $2t_{\eta}$; in terms of $t_e$, which is the equivalent timescale that is based on the turbulent magnetic diffusivity (which is probably the most relevant timescale for these dynamos), the most subcritical cases are run for longer than $20t_e$. 

We reproduced the subcritical dynamo simulation of SH04 over a running time of $1.64{\tiny }t_{\eta}$ or $21 t_e$. The simulation was initialised by a snapshot from $\Ra=1.18\Ra_c$ (Case I1). Pushing the Rayleigh number towards smaller values, SH04 found that the dynamo did not survive for $\Ra/\Ra_c=0.89$ (a result that we also recover, Case I5). This suggests that the subcriticality remains confined to a small range of $\Ra$ near $\Ra_c$. 

\begin{figure}
	\centering
	\subfigure[]{\label{fig:G5_meanfield_ts}
		\includegraphics[height=5cm]{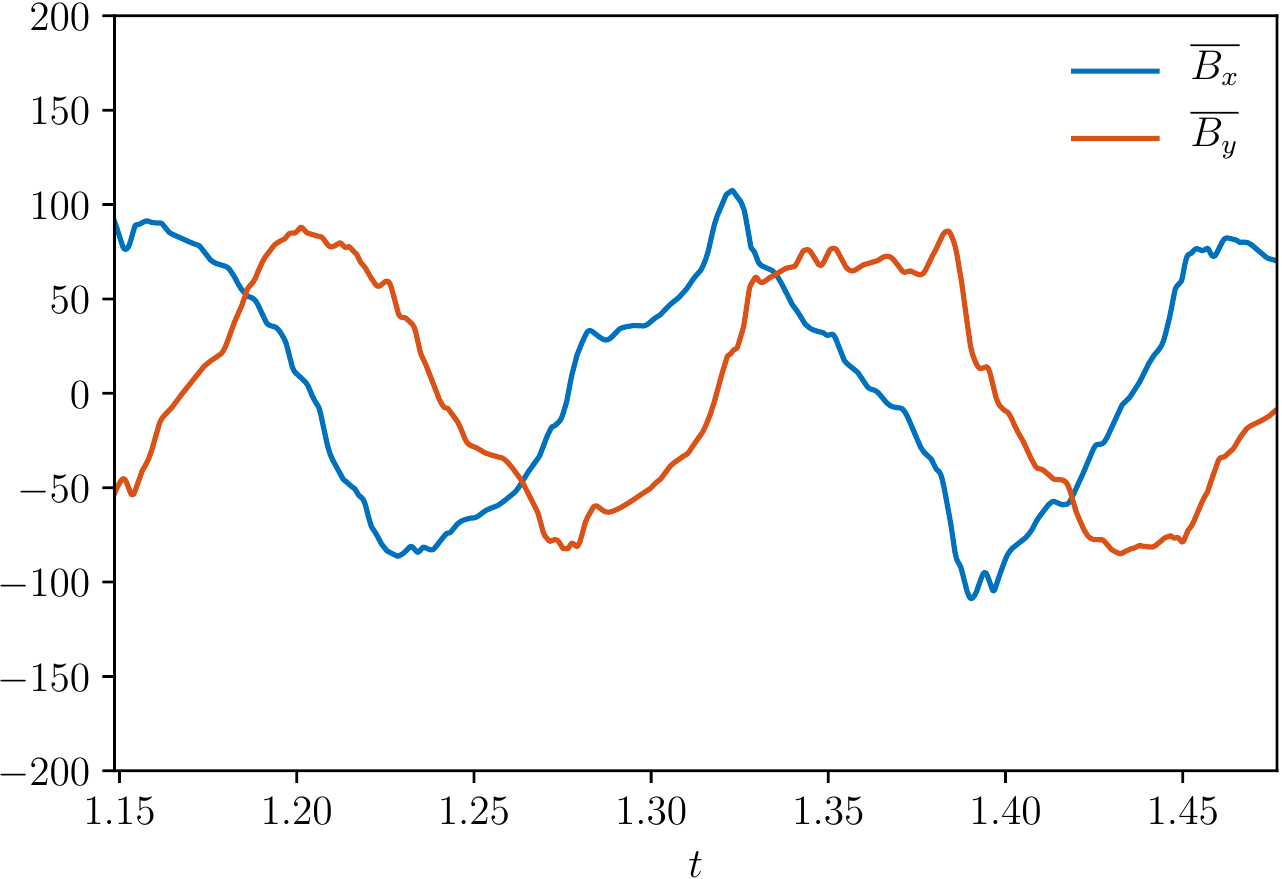}}
	\subfigure[]{\raisebox{2.8mm}{\label{fig:G5_meanfield}
			\includegraphics[height=4.75cm]{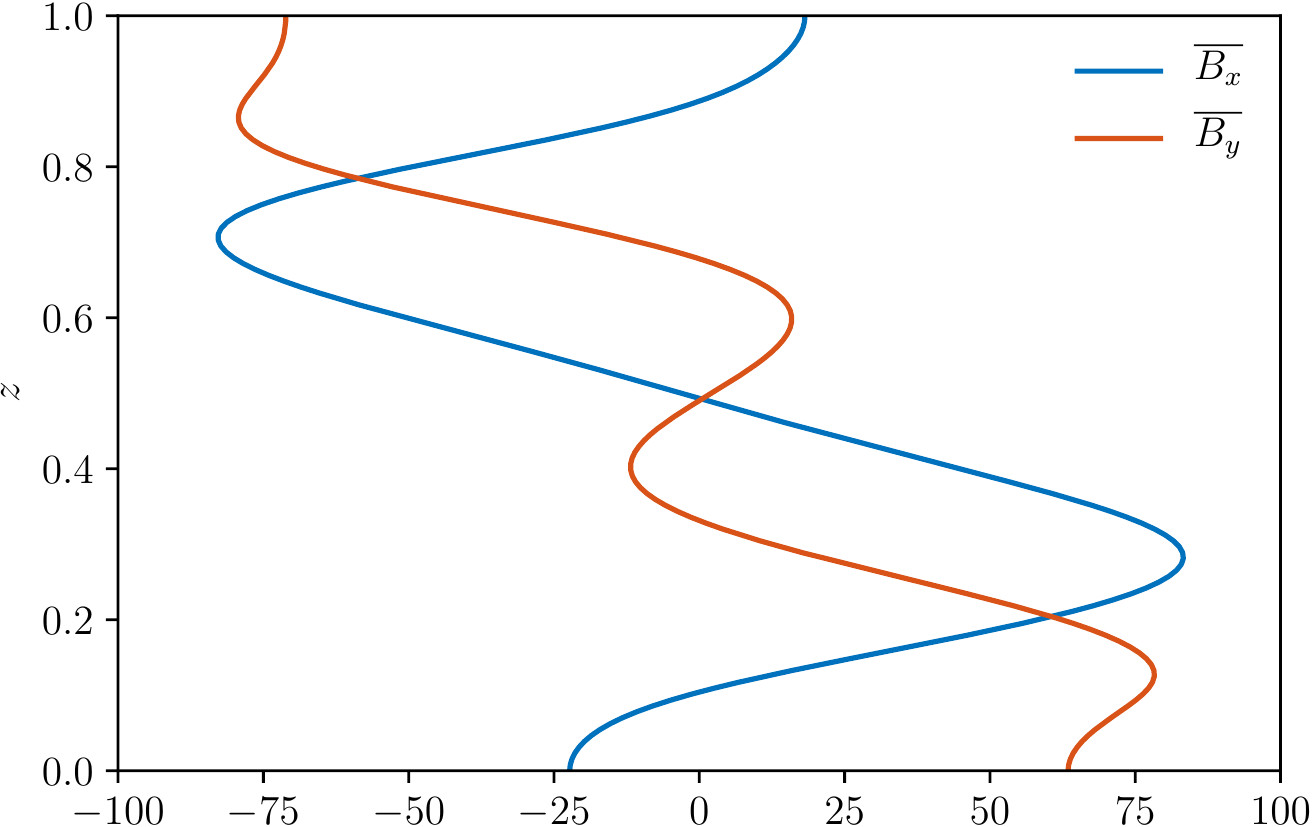}}}
	\caption{(a) Time series of the $x$ and $y$ components of the mean magnetic field at $z=0.85$ and (b) snapshot of its vertical profile in the Case G5 at $\Ra/\Ra_c=0.91$.}
	\label{fig:B_subcrit}
\end{figure}

Similarly to the supercritical case, the horizontally-averaged mean field oscillates over time as shown in Figure~\ref{fig:G5_meanfield_ts}. However, in the subcritical regime the time period of oscillation tends to be larger in comparison to the supercritical time period. The mean field is therefore oscillating more slowly below the onset of convection. Figure~\ref{fig:G5_meanfield} shows that the mid-plane antisymmetry, phase lag and spiral staircase structure observed in the supercritical case and SH04 remain. The dominant force balances observed are also similar to those in the supercritical saturated phase. At least in qualitative terms, these subcritical dynamos are very similar to their supercritical counterparts in terms of their behaviour.

\begin{figure}
\centering
\subfigure[]{
\includegraphics[width=0.48\textwidth]{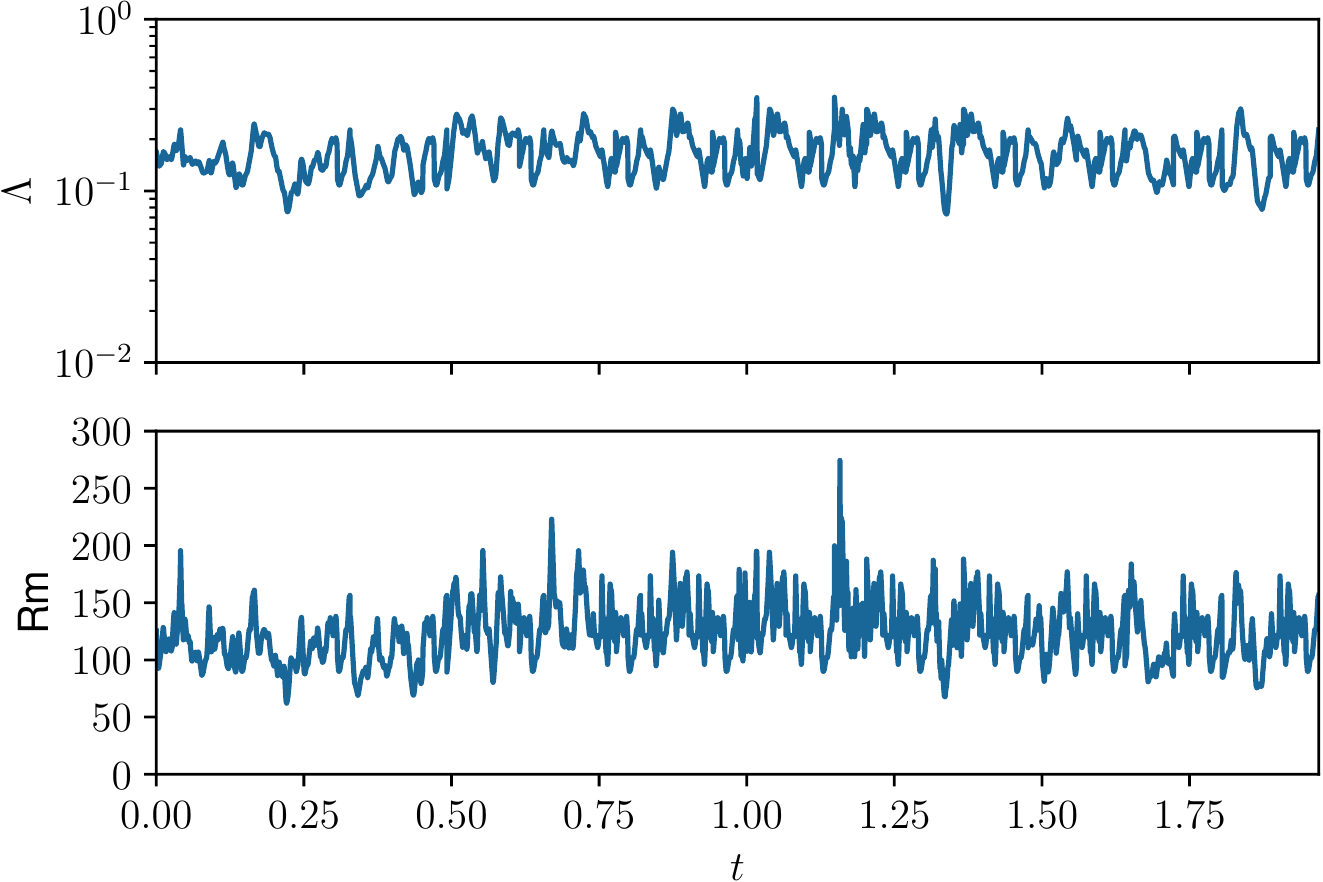}}
\subfigure[]{
\includegraphics[width=0.48\textwidth]{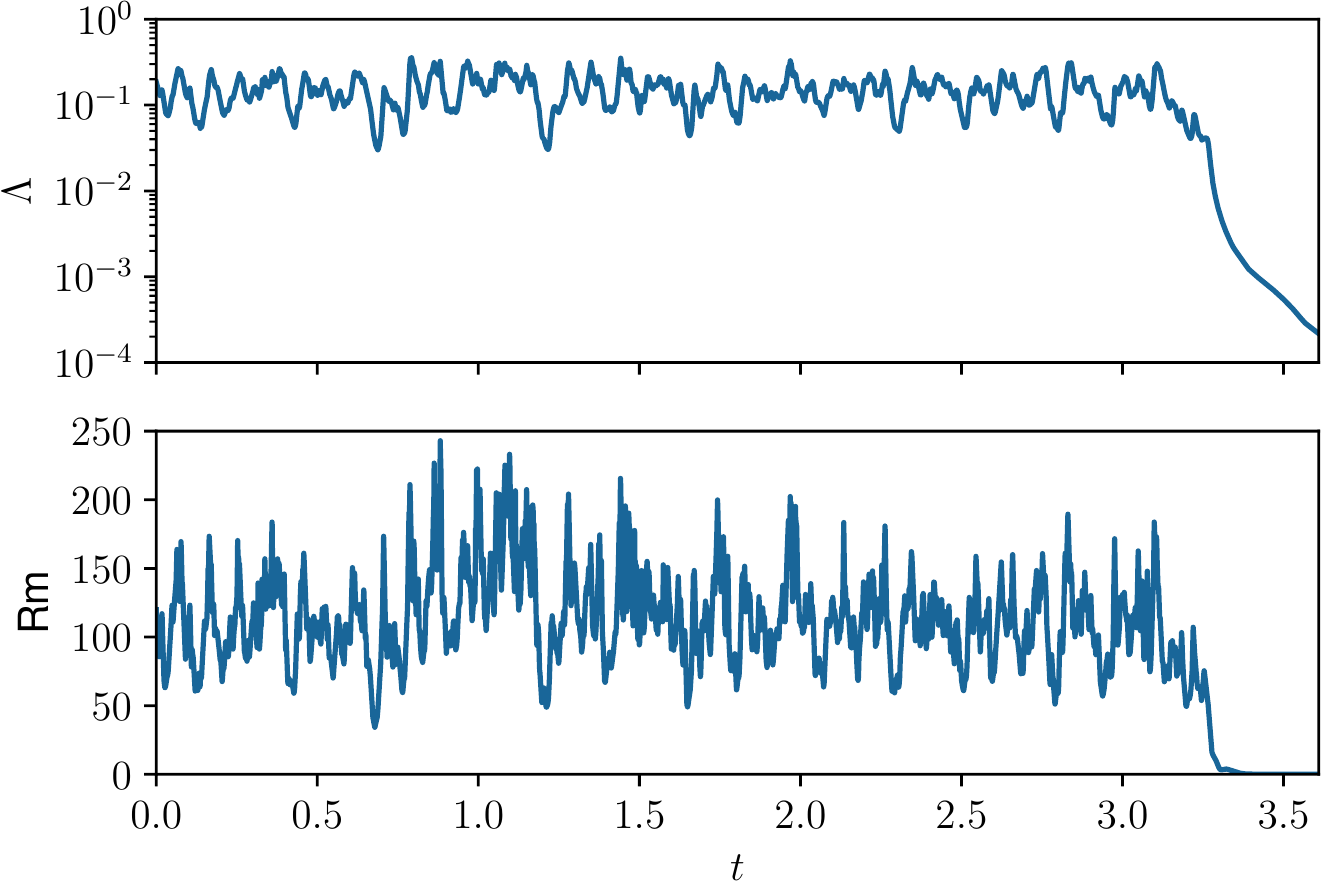}}
\caption{Time series of the Elsasser number, $\Lambda$, and magnetic Reynolds number, $\Rm$, for (a) Case H5 ($\Ra = 0.89 \Ra_c$) and (b) H6 ($\Ra= 0.87 \Ra_c$).}
\label{fig:Run6_ts}
\end{figure}

We explore the parameter space where the subcritical behaviour occurs to study the dependence on the Ekman number, which is varied between $3.16\times10^{-4}$ and $5\times10^{-6}$ for these simulations. For a given series of simulations at fixed $\Ek$ and $\Pm$, the subcritical behaviour is studied by branch tracking, where the initial condition at a given $\Ra$ is taken from a data snapshot at a slightly larger $\Ra$. 
Figure~\ref{fig:Run6_ts} shows the time series of $\Rm$ and $\Lambda$ for two representative cases: Case H5, where the dynamo is sustained for the whole duration of the simulation (which lasts $1.31 t_{\eta}$), and Case H6, where the magnetic field eventually decays. The collapse in this case occurs suddenly, after the magnetic field was sustained for approximately $1.34 t_{\eta}$. This collapse is likely due to a strong fluctuation that caused the system to transition to the non-dynamo (and non-convective) state. Pinpointing precisely the end point of the subcritical branch (\ie the lowest $\Ra/\Ra_c$ where dynamo occurs) would require multiple numerical realisations at the same parameters to examine the transition statistics \citep[\eg][]{VanKan19}. This technique would have a considerable computational cost for the resolutions required here. Consequently, we are only able to determine the approximate end point of each branch, with a possible dependence of their location on the running times of the simulations and on the aspect ratio of the computational domain. Indeed, the system might be more sensitive to fluctuations, and hence prone to a dynamo collapse, when the scale separation between the box size and horizontal convective scale is small. In order to test the impact of smaller aspect ratio, we reproduced Cases I1-I5 at $\Ek=5\times 10^{-6}$ and $\Pm=1$, reducing the computational domain to $\lambda=0.5$ and again to $\lambda=0.25$ with the horizontal resolution accordingly rescaled to $64^2$ and $32^2$ respectively and with similar run time. At $\lambda=0.5$ the dynamo behaved similarly, with global output quantities approximately unchanged on average but experiencing larger fluctuations. Subcritical dynamo action was sustained as far as $0.93\Ra_c$. At $\lambda=0.25$ fluctuations in the global output quantities were severe and resulted in the dynamo decaying at $0.98\Ra_c$. For this set of parameters, reducing the aspect ratio has therefore decreased the depth of the subcritical branch.

\begin{figure}
\centering
	\subfigure[]{\label{fig:Rm_Ra}
	\includegraphics[width=0.48\textwidth]{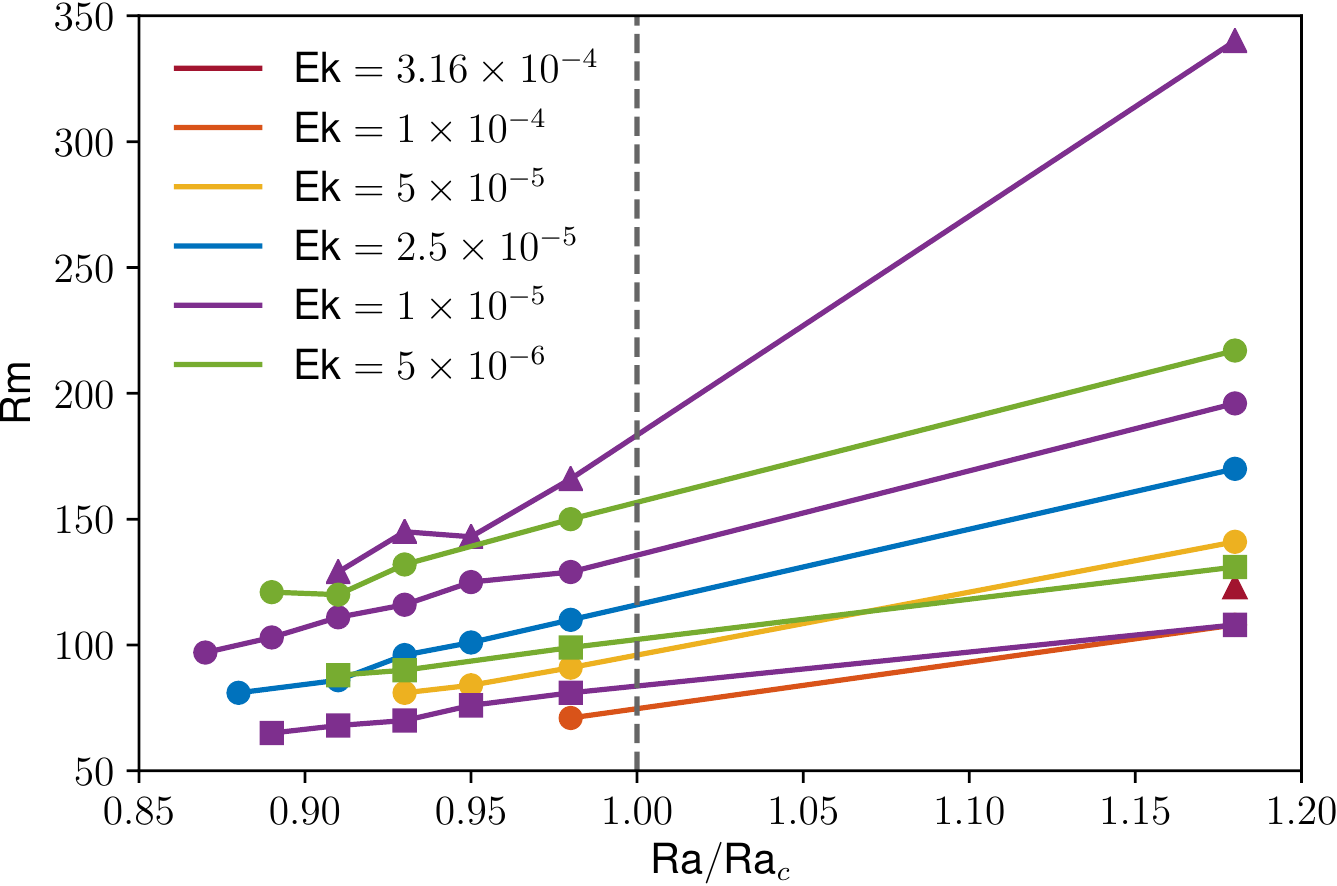}}
	\subfigure[]{\label{fig:Els_Ra}
	\includegraphics[width=0.48\textwidth]{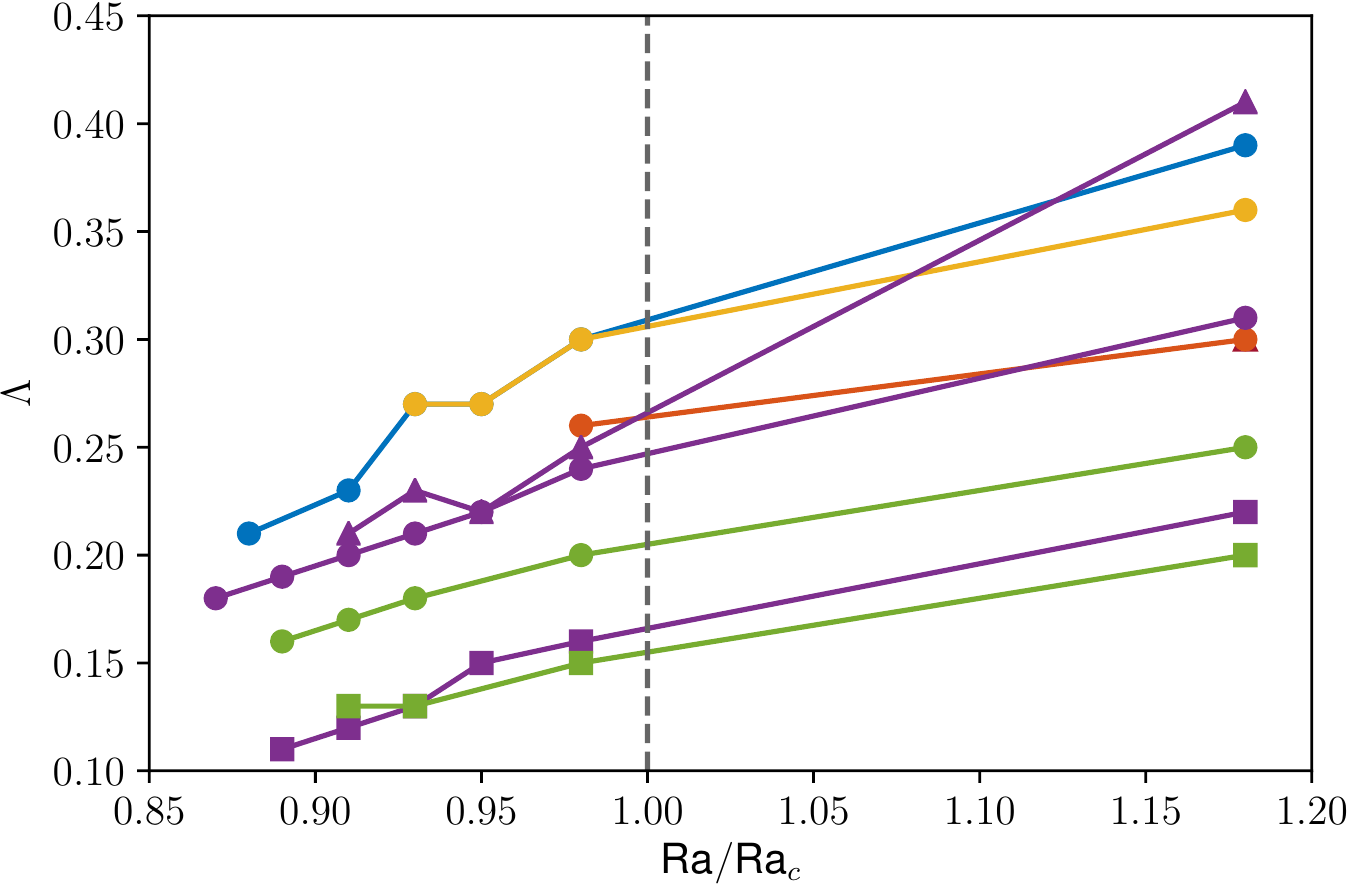}}
	\subfigure[]{\label{fig:ElsEk_Ra}
	\includegraphics[width=0.48\textwidth]{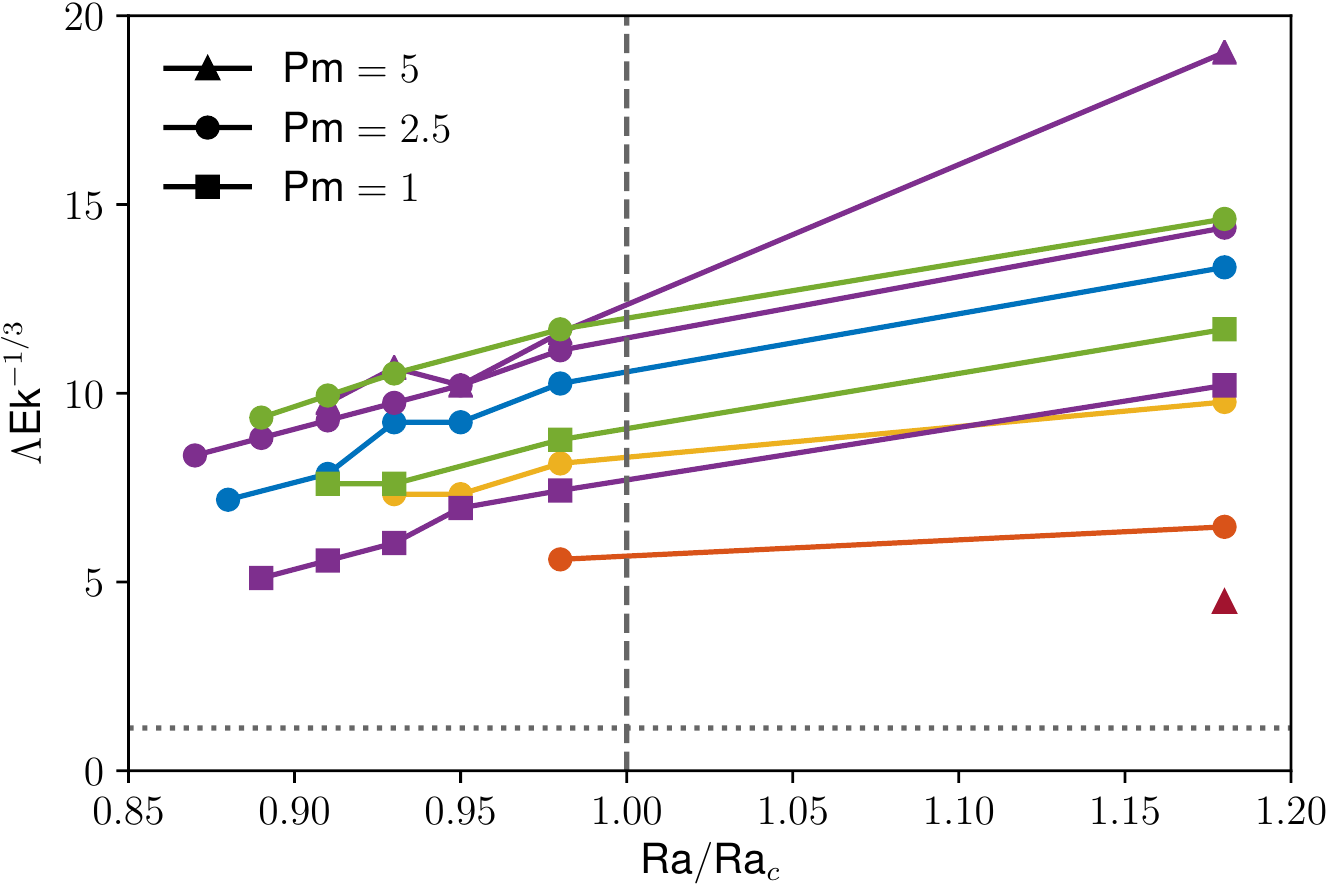}}
	\subfigure[]{\label{fig:Elsmean_Ra}
	\includegraphics[width=0.48\textwidth]{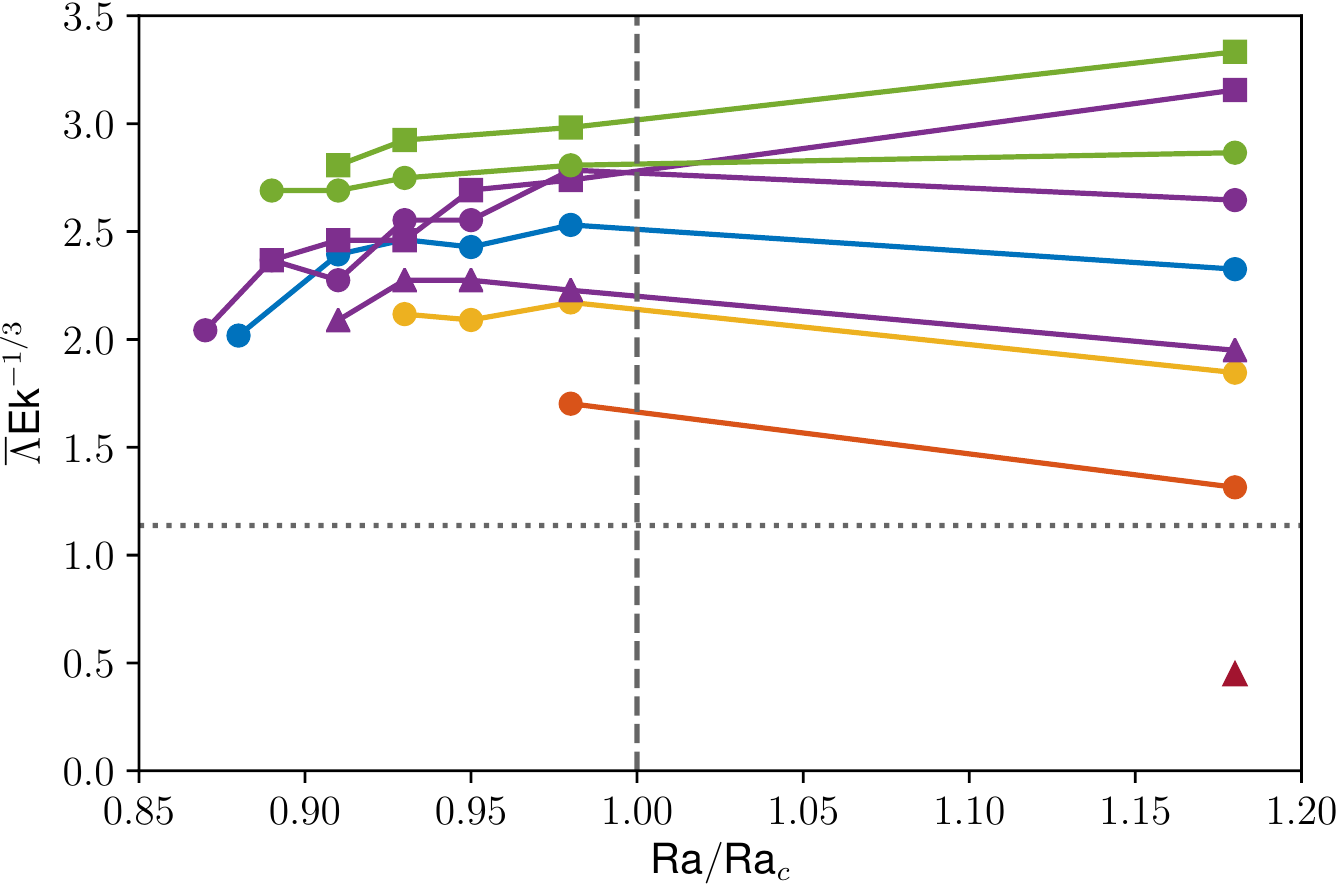}}
	\caption{The dependence of some of the global output quantities as a function of $\Ra/\Ra_c$: (a) $\Rm$, (b) $\Lambda$, (c) $\Lambda \Ek^{-1/3}$ and (d) $\overline{\Lambda}  \Ek^{-1/3}$. The horizontal dotted line in (c)-(d) represents the switch-over value $\Lambda_{sw} \Ek^{-1/3} = 1.14$. As in Figure~\ref{fig:runs}, each colour corresponds to a different value of $\Ek$, as indicated in (a), whilst the different symbols correspond to different choices of $\Pm$, as indicated in (c).}
\label{fig:RmEls_Ra}
\end{figure}

Figure~\ref{fig:runs} shows the location of the failed and sustained dynamo simulations for each series. The range of $\Ra$ where subcritical dynamos are maintained deepens when $\Ek$ decreases in the moderate Ekman number regime. The effect of increasing $\Pm$ is not monotonic: the subcritical range deepens when $\Pm$ increases from $1$ to $2.5$, but shrinks when $\Pm$ is increased further (the generally stronger fluctuations at higher $\Pm$ are more likely to cause the solution to drop off the dynamo branch). 
To understand the dependence of the subcritical range on $\Ek$ and $\Pm$, we consider the necessary conditions for the existence of subcritical dynamos.
At least two conditions must be met: (i) a dynamo producing a coherent mean field must be operating, and (ii) the mean magnetic field must be of sufficiently large amplitude to alter the convective flows. The first condition is quantified by the existence of a threshold magnetic Reynolds number, $\Rm_t$, and the second condition by a threshold Elsasser number, $\Lambda_t$. A dynamo can only survive in the subcritical range if $\Rm>\Rm_t$ and $\Lambda>\Lambda_t$. Violating one of these conditions would lead to a dynamo collapse and can possibly mark the location of the end point of the subcritical branch. 
In practice, determining whether the transition $\Rm\approx\Rm_t$ or $\Lambda\approx\Lambda_t$ corresponds to the end points might prove difficult if the threshold values depend on $\Ek$ and $\Pm$ and because of the approximate location of the end points from the numerical results. Determining a precise threshold at which dynamo action can no longer be sustained is clearly difficult as we cannot accurately determine whether a fluctuation is sufficiently strong enough to cause the dynamo to decay, especially considering that fluctuations are local events which may not accurately be captured by a globally averaged quantity such as $\Lambda$ or $\Rm$.

With these caveats in mind, we plot $\Rm$ as a function of $\Ra/\Ra_c$ in Figure~\ref{fig:Rm_Ra}. The end points of each series are located around $\Rm\approx70-100$, although some dispersion exists due to the large temporal fluctuations of $\Rm$.  Overall, our simulations are in broad agreement with the idea that the dynamo collapse occurs when $\Rm\approx \Rm_t$. For a given $\Pm$ at fixed $\Ra/\Ra_c$, $\Rm$ increases when $\Ek$ decreases, which is consistent with the deepening of the subcritical range in this case. For fixed $\Ek$ and $\Ra/\Ra_c$, increasing $\Pm$ leads to a increase of the mean value of $\Rm$, which is favourable for the subcriticality up to a point, but it also leads to an increase of the temporal fluctuations of $\Rm$ (see the standard deviation values in Table~\ref{table1} for the series E, F and G for instance), which is detrimental for subcritical dynamo action because of the higher chances of pushing the system towards the non-dynamo state. These two competing effects lead to an optimal value of $\Pm$, which here is around $\Pm\approx2.5$.  

Figure~\ref{fig:Els_Ra} shows $\Lambda$ as a function of $\Ra/\Ra_c$. For $\Ek<10^{-4}$, $\Lambda$ decreases with $\Ek$ for a fixed $\Ra/\Ra_c$ and $\Pm$.  The subcritical branch terminates when $\Lambda<0.27$ for $\Ek=5\times10^{-5}$ and when $\Lambda<0.17$ for $\Ek=5\times10^{-6}$ at $\Pm=2.5$, so the dependence of the threshold value of $\Lambda$ is weakly dependent on $\Ek$.
\citet{Elt72} studied linear magnetoconvection in the presence of rotation (with the rotation axis aligned with the direction of gravity) and an imposed uniform horizontal magnetic field. Similarly to the case with imposed uniform vertical field \citep{Chandra61}, three dynamically distinct regimes exist at a given Ekman number depending on the value of the Elsasser number that controls the strength of the imposed field, $\Lambda_0$: a rotation-dominated regime for $\Lambda_0<\mathcal{O}(\Ek^{1/3})$, a magnetically-dominated regime for $\Lambda_0>\mathcal{O}(1)$, and an intermediate regime, where Coriolis and Lorentz forces are of equal importance. The critical Rayleigh number in the intermediate regime is always smaller than in the rotation-dominated regime for low $\Ek$, so the presence of an imposed horizontal field favours convection in this regime. The switch-over value of $\Lambda_0$ that marks the transition from the rotation-dominated regime and the intermediate regime occurs for $\Lambda_{sw}= 1.14 \Ek^{1/3}$ for the same boundary conditions as the ones considered here (Equation~\eqref{eq:BC}). Although the mean field generated in our fully nonlinear dynamo simulations is not uniform, $\Lambda_{sw}$ could be regarded as a lower bound for the threshold value $\Lambda_t$ required for the existence of magnetically-modified convection and subcritical dynamos.  
Figure~\ref{fig:ElsEk_Ra} shows $\Lambda\Ek^{-1/3}$ as a function of $\Ra/\Ra_c$ in our simulations.  The subcritical branches terminate for $\Lambda \Ek^{-1/3}\approx 8-10$ when $\Ek<10^{-4}$ and are always located above the theoretical switch-over value $\Lambda_{sw} \Ek^{-1/3} = 1.14$. This is consistent with $\Lambda_{sw}$ being a lower bound for $\Lambda_t$. Since it is mainly the mean magnetic field that alters the convection, considering $\overline{\Lambda} \Ek^{-1/3}$, where $\overline{\Lambda}$ is based on the mean field amplitude only, might be more suitable. The evolution of $\overline{\Lambda} \Ek^{-1/3}$ as a function of $\Ra/\Ra_c$ plotted in Figure~\ref{fig:Elsmean_Ra} shows that $\overline{\Lambda} \Ek^{-1/3}$ takes values around $2-4$ in the subcritical range and is, again, always above the theoretical switch-over value. The only Ekman number for which we did not observe a subcritical behaviour ($\Ek=3.16\times10^{-4}$) has a supercritical dynamo which produces values of $\overline{\Lambda} \Ek^{-1/3}$ below the theoretical switch-over value.

\begin{figure}
\centering
\includegraphics[width=0.5\textwidth]{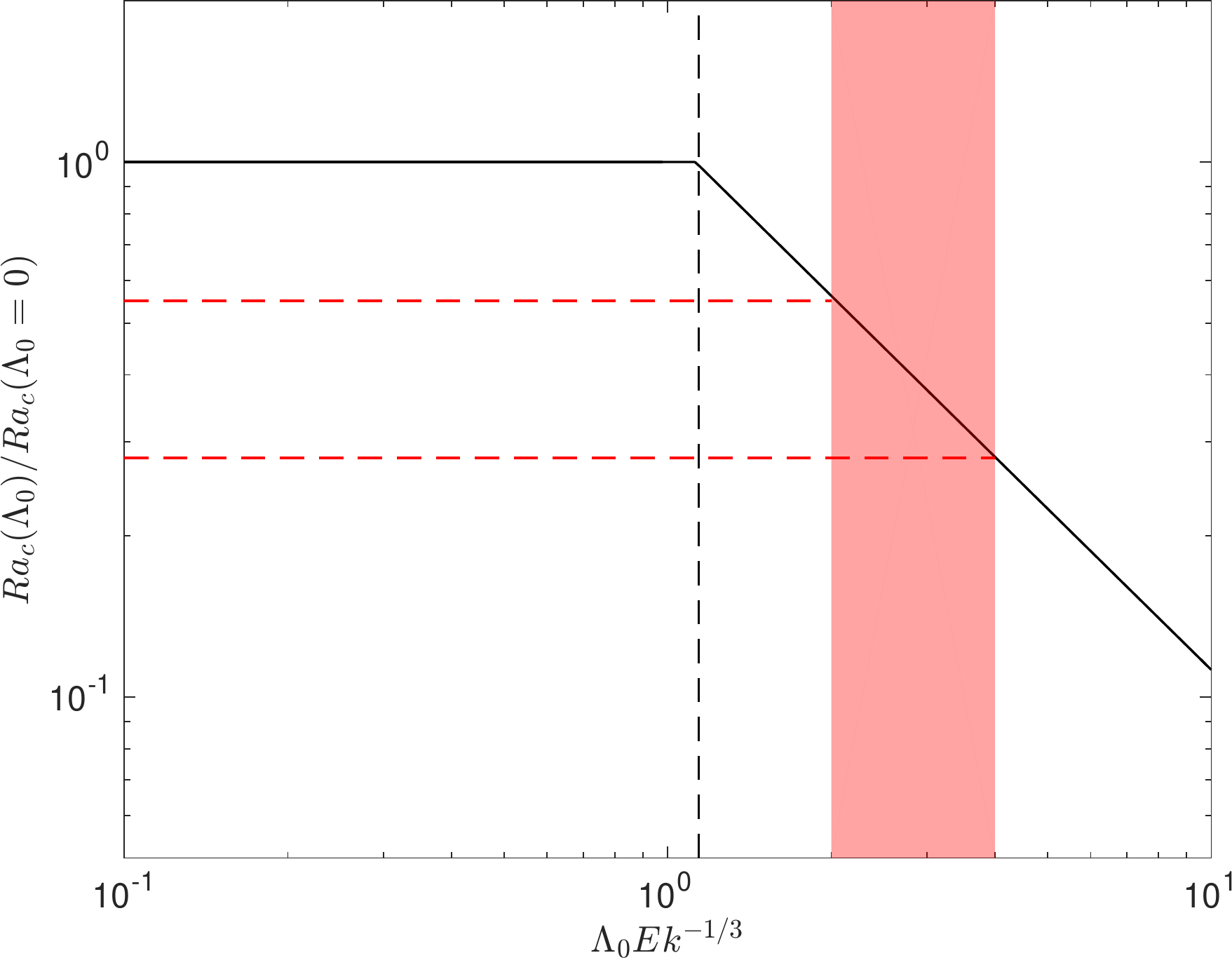} 
\caption{Critical Rayleigh number at the linear onset of rotating magnetoconvection \citep[from][]{Elt72} normalised by the critical Rayleigh number in the non-magnetic case as a function of $\Lambda_0 \Ek^{-1/3}$, where $\Lambda_0$ is the Elsasser number based on the imposed field strength, in the case where the rotation axis is aligned with gravity and the imposed field is uniform and horizontal. The boundary conditions are the same as in equation~\eqref{eq:BC}. The vertical dashed line corresponds to the switch-over value $\Lambda_{sw} \Ek^{-1/3}= 1.14$. The red area corresponds to the approximate values of $\overline{\Lambda} \Ek^{-1/3}$ in our subcritical simulations.}
\label{fig:onset_theo}
\end{figure}

If the magnetoconvection set-up of \citet{Elt72} is a suitable simplified model to explain the results of our nonlinear simulations near the onset of convection, then a third condition exists for the depth of the subcritical range at fixed $\Ek$: $\Ra$ must be greater than the theoretical critical Rayleigh number for convective onset at a given $\Lambda$, denoted by $\Ra_c(\Lambda)$.  $\Ra_c(\Lambda)$ is the smallest possible $\Ra$ where subcritical dynamos might occur. Figure~\ref{fig:onset_theo} shows $\Ra_c(\Lambda_0)$ normalised by $\Ra_c(\Lambda_0=0)$ as a function of $\Lambda_0 \Ek^{-1/3}$ in the rotation-dominated regime ($\Lambda_0<\Lambda_{sw}$) and in the intermediate regime ($\Lambda_0>\Lambda_{sw}$) based on the results of \citet{Elt72} in the same configuration as described earlier. The minimum values of $\overline{\Lambda} \Ek^{-1/3}$ obtained in our subcritical dynamos are around $2-4$, which corresponds to values of $\Ra_c(\Lambda_0)/\Ra_c(\Lambda_0=0)$ around $0.3-0.6$. The subcritical range found in our simulations, which is always above $\Ra=0.85\Ra_c(\Lambda_0=0)$, is consistent with these theoretical lower values. The quantity $\overline{\Lambda} \Ek^{-1/3}$ tends to increase when $\Ek$ decreases (for a fixed $\Pm$), although this dependence is relatively small. This suggests that the possible subcritical range might extend slowly below $0.3-0.6\Ra_c(\Lambda_0=0)$ for dynamos at lower $\Ek$ and $\Pm=\mathcal{O}(1)$.

To conclude this section, we study the role of the kinetic helicity in our subcritical dynamos.
In a study of dynamos driven by spherical rotating convection, \citet{Sree11} found that dipolar magnetic fields feedback on the flow such that they enhance the kinetic helicity. As kinetic helicity is often thought to be an important ingredient for large-scale dynamo action \citep[e.g.][]{Moffatt78}, its enhancement might lead to the subcritical behaviour observed in the simulations of \citeauthor{Sree11}. To test whether a similar idea applies in our planar dynamos, we have compared the vertical profiles of the horizontally-averaged kinetic helicity $\mathcal{H}$ in the kinematic and saturated phases of the supercritical Case I1 (not shown). The comparison of the two phases is not entirely straightforward because the flow is more vigorous in the saturated phase, so we have also compared the profiles of the relative kinetic helicity $\mathcal{H}_{rel}$.  The profiles are approximately sinusoidal (antisymmetric about the mid-plane), which is typical of planar rotating convection at low $\Ra$ \citep[\eg][]{Cat06}. The maxima of $\mathcal{H}$ increase by approximately 20\% in the saturated phase, but the maxima of $\mathcal{H}_{rel}$ decrease by approximately 30\%. The increase of $\mathcal{H}$ that we observe here is very small compared with the increase measured by \citeauthor{Sree11} in the presence of a dipolar magnetic field, suggesting that the enhancement of the helicity by the action of the magnetic field is probably not responsible for subcritical behaviour in our system. 

\subsection{Intermittent dynamos at small Ekman numbers}
\label{sec:E5e-7}

\begin{figure*}
	\centering
	\subfigure[]{\label{fig:J1_ts}
		\includegraphics[width=0.48\textwidth]{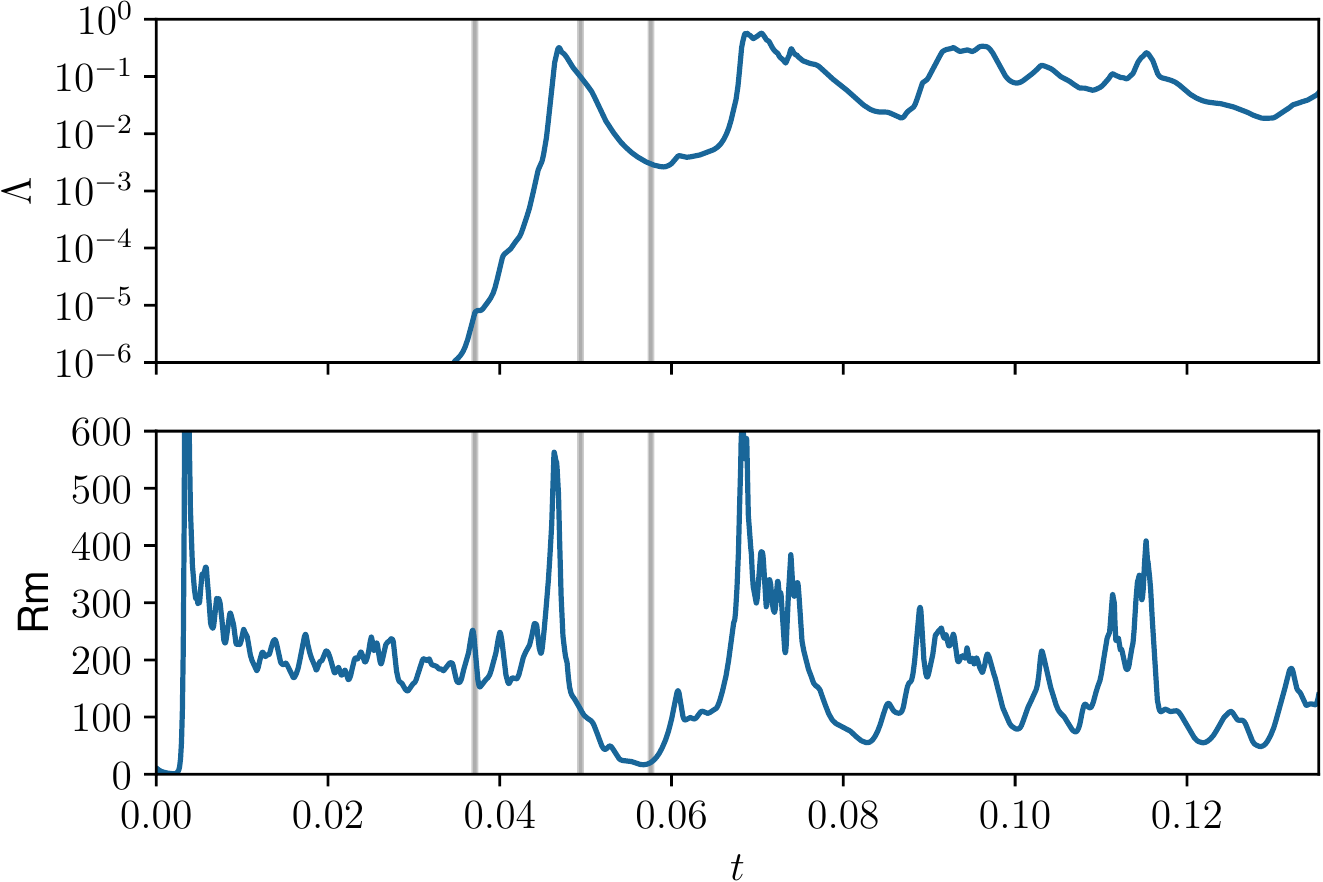}}
		\subfigure[]{
		\includegraphics[width=0.48\textwidth]{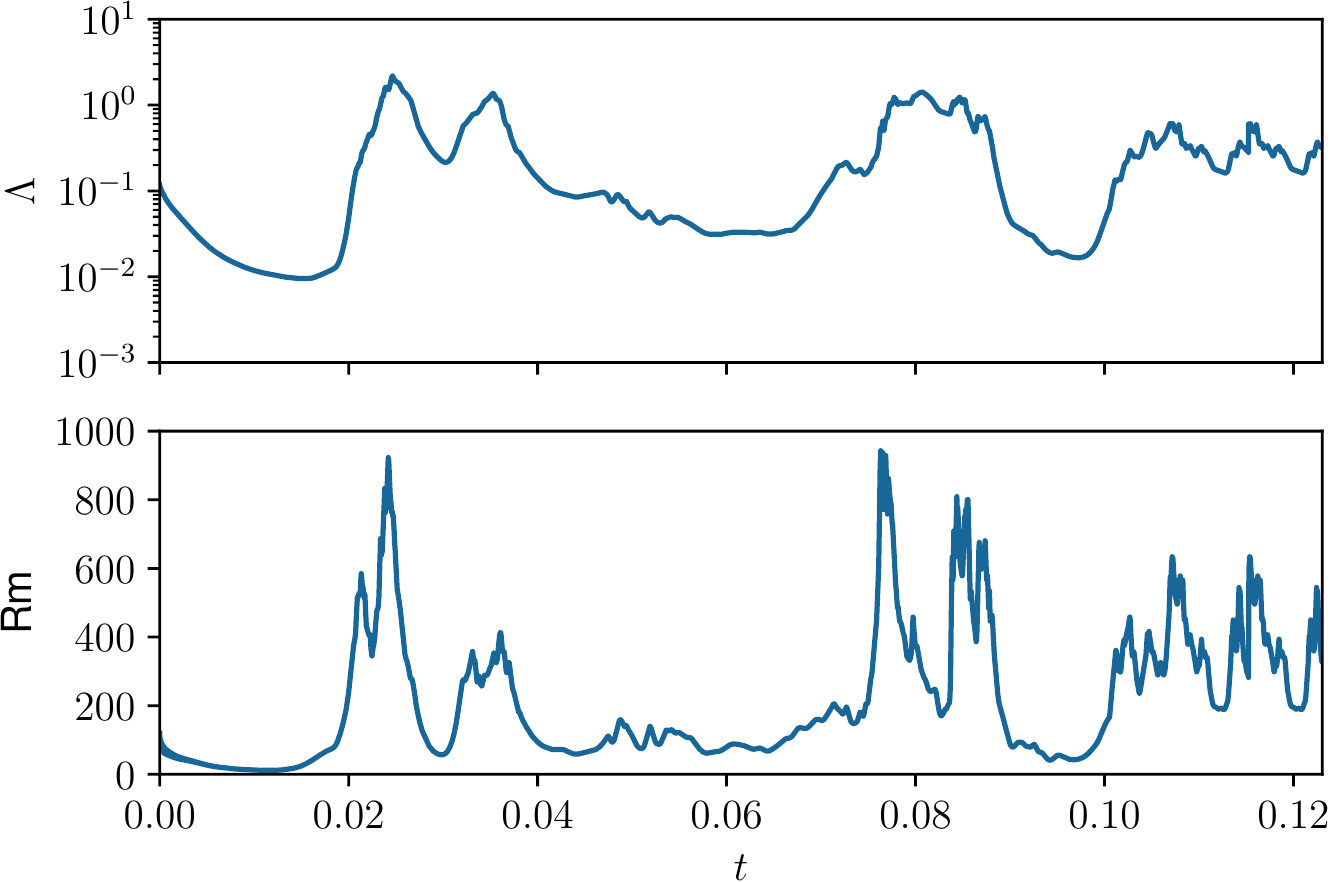}}
	\caption{Time series of the Elsasser number $\Lambda$ and magnetic Reynolds number $\Rm$ for Case J1 ($\Ra/\Ra_c=1.18$) and Case J2 ($\Ra/\Ra_c=0.98$). The grey lines indicate the locations of the snapshots presented below.}
	\label{fig:Ek5e-7_ts}
\end{figure*}

At $\Ek=5\times10^{-7}$, the behaviour of the supercritical dynamo in the saturated phase changes, as shown in Figure~\ref{fig:Ek5e-7_ts}(a), which shows the time series of $\Lambda$ and $\Rm$ in Case J1 ($\lambda=0.25$) for $\Ra/\Ra_c=1.18$ and $\Pm=1$. Note that for this low Ekman number, the computational domain is restricted to small aspect ratios to ensure that the convective motions can be resolved without using substantially more grid points for the simulations. For $\lambda=0.25$, there are approximately seven convective cells across the domain in the purely hydrodynamical case near the convective onset, so the scale separation between the box and convective scale is modest, but still probably sufficient to study large-scale dynamo action \citep{HC_2008}. The presence of large-scale motions may however result in some box-sized dependence on the dynamo. It is also possible that subcritical dynamos in smaller domains will be more prone to disruption by fluctuations at lower $\Ek$. However, this reduction in domain size is a pragmatic choice that allows the lower $\Ek$ regime to be explored.

\begin{figure}
	\centering
		\subfigure[]{
		\includegraphics[width=0.32\textwidth]{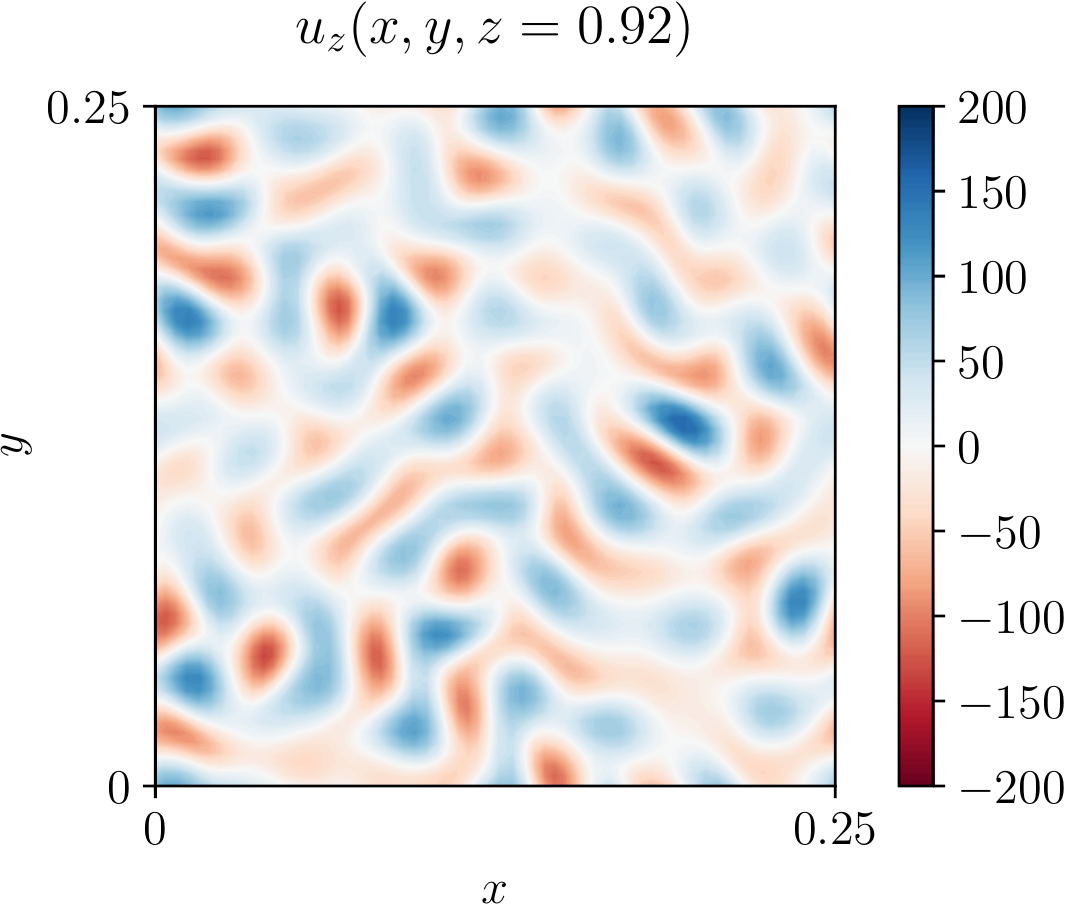}}
			\subfigure[]{
		\includegraphics[width=0.32\textwidth]{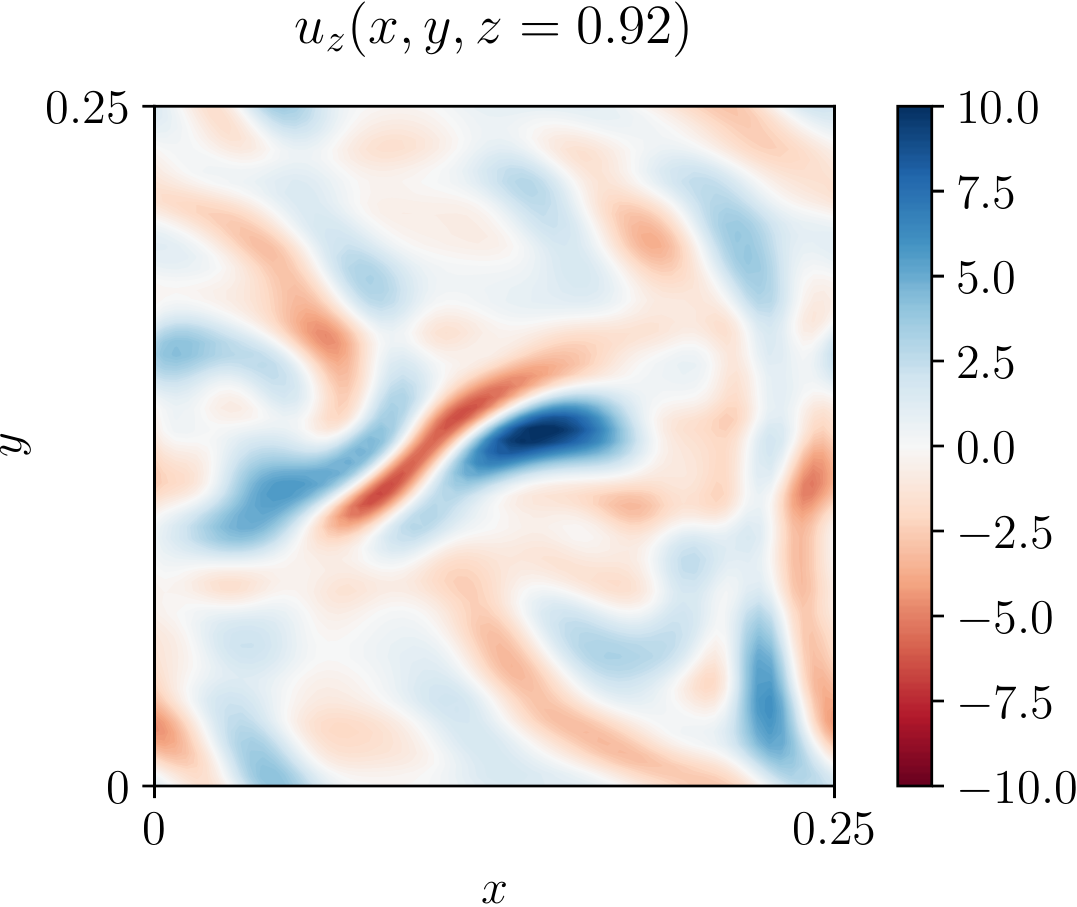}}
			\subfigure[]{
		\includegraphics[width=0.32\textwidth]{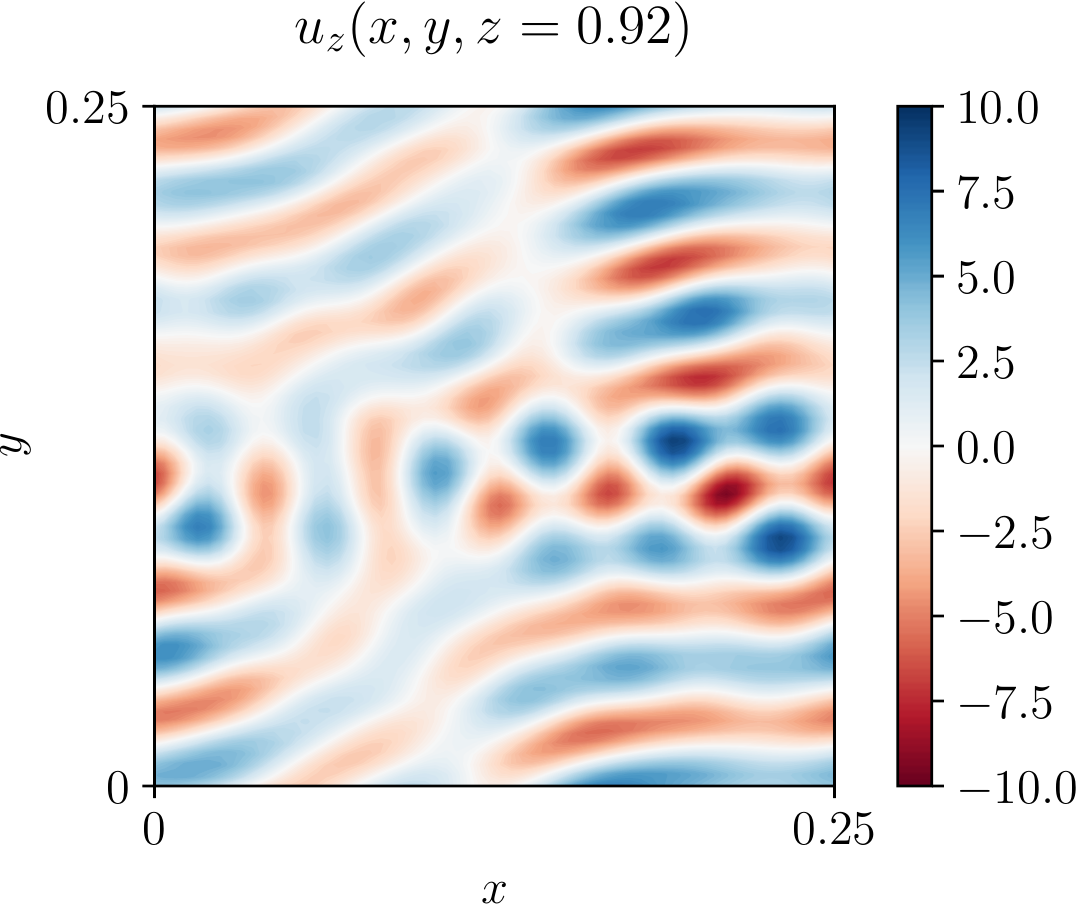}}
	\subfigure[]{
	\includegraphics[width=0.32\textwidth]{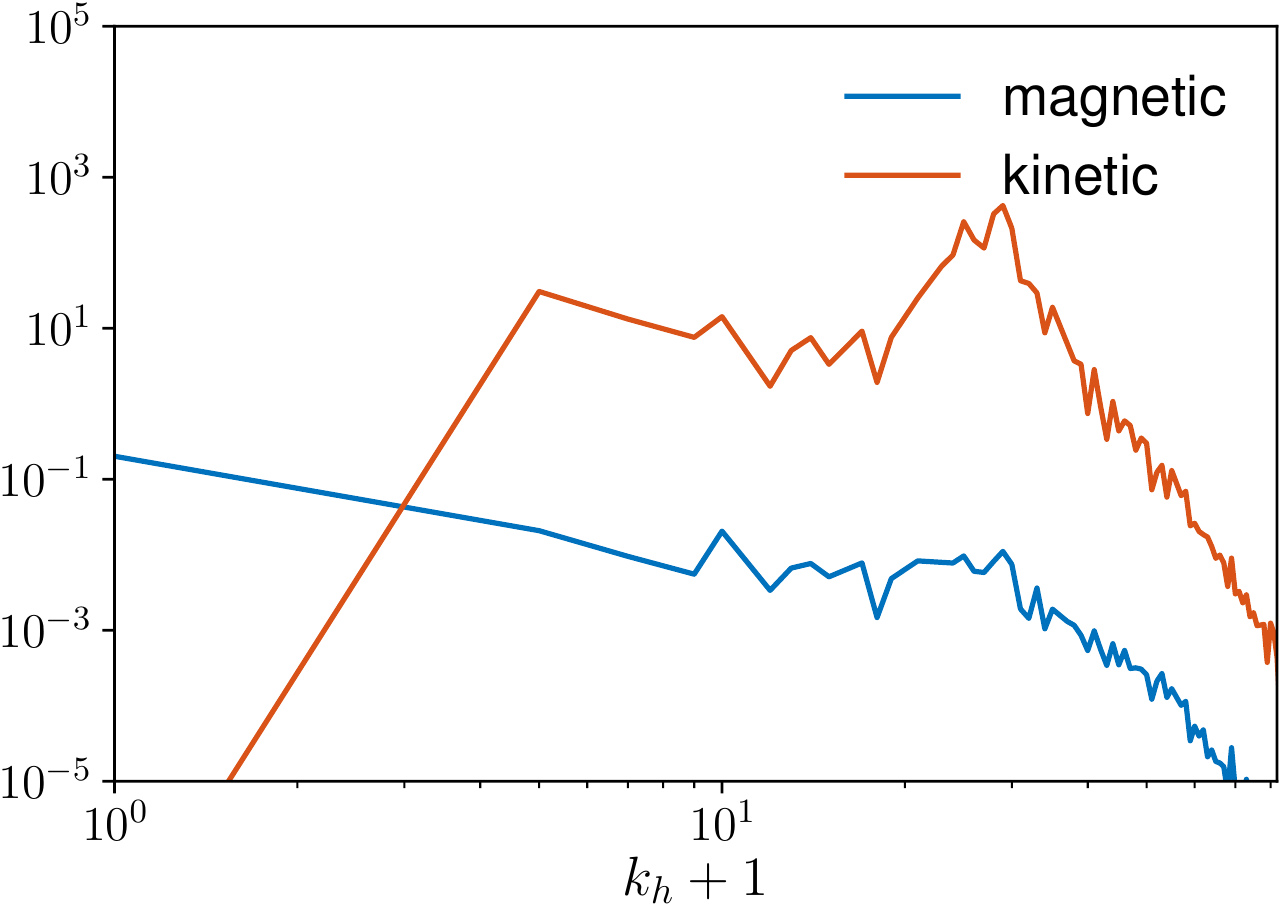}}
\subfigure[]{
	\includegraphics[width=0.32\textwidth]{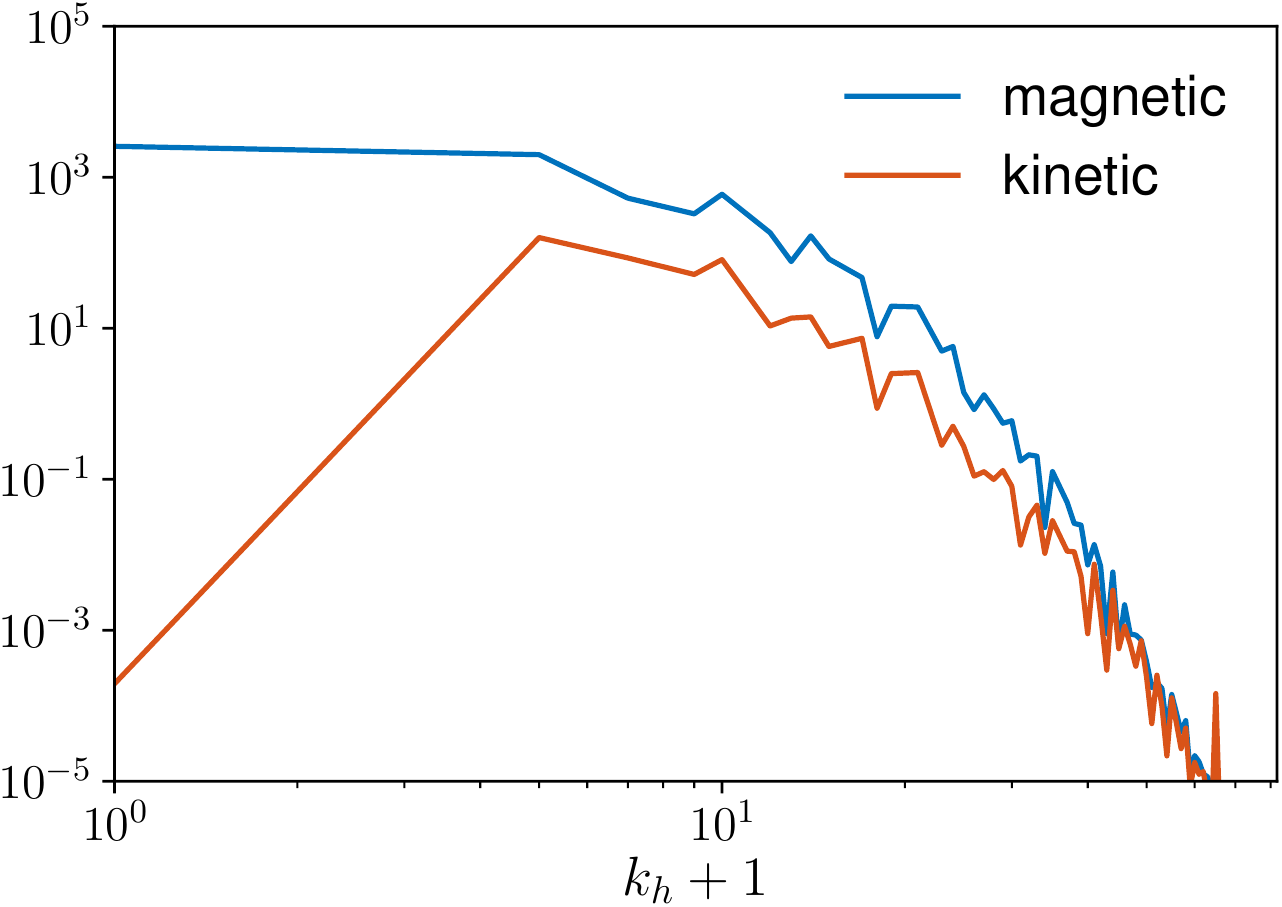}}
\subfigure[]{
	\includegraphics[width=0.32\textwidth]{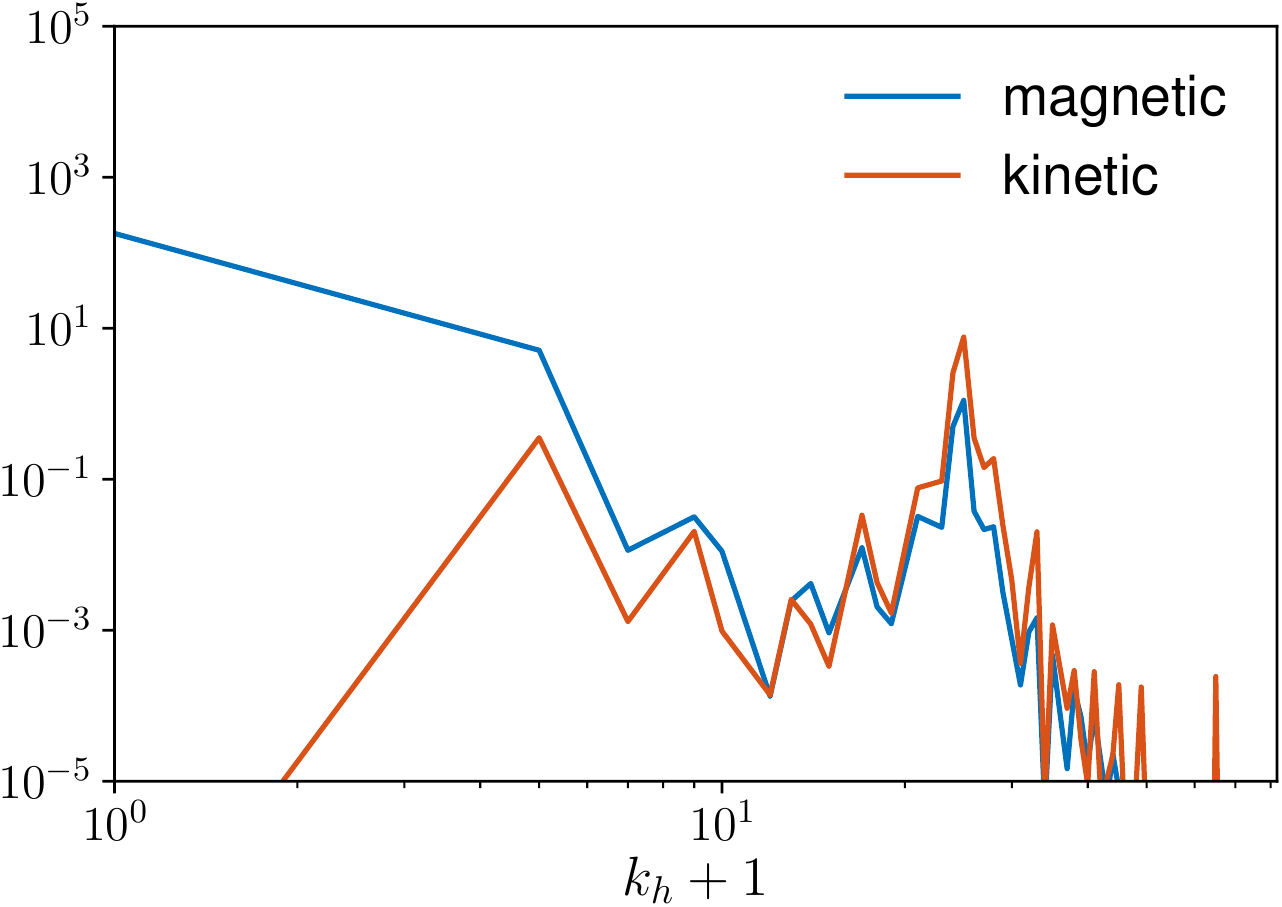}}
	\caption{Snapshots of the horizontal distribution of the vertical velocity (at $z=0.92$) for Case J1 in (a) the kinematic phase, (b) near the peak of $\Lambda$ and (c) during the decaying phase. The bottom row shows the corresponding power spectra for the magnetic energy (blue) and the kinetic energy (red), as a function of the horizontal wavenumber for (d) the kinematic phase, (e) near the peak in $\Lambda$ and (f) during the decaying phase. The locations of each of these snapshots are represented by grey lines in Figure~\ref{fig:J1_ts}. The box-size mode corresponds to $k_h=4$ for this aspect ratio.}
	\label{fig:Ek5e-7_0p5}
\end{figure}

In Case J1, we observe an intermittent behaviour: once the magnetic field reaches a sufficiently large amplitude, corresponding to $\Lambda\approx10^{-0.5}$, the kinetic energy decays under the influence of the magnetic field, which (in turn) leads to a decay of the magnetic energy. Once $\Lambda$ is sufficiently small, of the order of $10^{-2.5}$, the kinetic energy starts to increase again as the convection recovers. This is followed by a subsequent growth of the magnetic energy and the cycle continues. SH04 demonstrated a similar intermittent behaviour for a slightly less supercritical case ($\Ra/\Ra_c=1.09$) for the same aspect ratio. This intermittency is due to drastic changes in the flow during the phases of larger ($\Lambda\approx10^{-0.5}$) and weaker ($\Lambda\approx10^{-2.5}$) field strength. Horizontal slices of the vertical velocity at different times are shown in Figure~\ref{fig:Ek5e-7_0p5}. At times corresponding to high values of $\Lambda$, the flow takes the form of a (near) box-sized mode. This large-scale mode is associated with an enhanced vertical velocity and temperature perturbation, and is thus efficient at transporting heat across the layer (corresponding to an increase in the Nusselt number of $180\%$ compared with the purely hydrodynamical state).  The emergence of this large-scale convection mode is a much more dramatic indication of the dynamical importance of the Lorentz forces than the fat localised convection columns observed in Figure~\ref{fig:flow_sup} during the saturated phase of the dynamos at more moderate $\Ek$ (\S\ref{sec:super}). The collapse of the magnetic energy following the emergence of the large-scale mode indicates that this type of flow is inefficient at driving the dynamo necessary to maintain a sufficiently strong magnetic field. The kinetic energy spectra in Figure~\ref{fig:Ek5e-7_0p5} show that the small-scale convection is suppressed when the field is strong; as these scales of motion are the dominant contributors to the mean e.m.f., this inhibits the dynamo. As the magnetically-driven large-scale mode decays with the magnetic field, the small-scale convection re-emerges, which allows the dynamo to recover. 

Whilst the Elsasser number for given $\Ra/\Ra_c$ and $\Pm$ decreases with $\Ek$ in our (more moderately rotating) dynamo calculations (Figure~\ref{fig:Els_Ra}), the important parameter for the transition to magnetically-modified convection is $\overline{\Lambda}\Ek^{-1/3}$, which increases when $\Ek$ decreases (Figure~\ref{fig:ElsEk_Ra}).  For $\Ek=5\times10^{-7}$, $\overline{\Lambda}\Ek^{-1/3}\approx 10$ at the peak of the magnetic energy, so the system enters more deeply into the intermediate regime described earlier than the cases at moderate $\Ek$, which remain on its margins with $\overline{\Lambda} \Ek^{-1/3}\approx2-4$ (Figure~\ref{fig:onset_theo}).
The intermittent behaviour is therefore due to the ability of the small-scale convection to sustain magnetic fields of relatively stronger amplitude at low Ekman numbers, and then the apparent inability of the subsequent magnetically-driven large-scale convection mode to sustain the dynamo.
              
Figure~\ref{fig:Ek5e-7_ts}(b) shows the time series of $\Lambda$ and $\Rm$ in Case J2 ($\lambda=0.25$) for $\Ra/\Ra_c=0.98$ and $\Pm=1$. This is in the weakly subcritical regime, and has only been evolved for a short period of time. However, it exhibits similar intermittent behaviour to its supercritical counterpart. This exploration of the subcritical regime will be continued in future work, but these preliminary results (albeit at low aspect ratio) suggest that it should still be possible to find convincing candidates for subcritical dynamo action in this low $\Ek$ regime, at least over a limited range of Rayleigh numbers (Figure~\ref{fig:runs}). 

\section{Conclusion}
\label{sec:discussion}

We have studied dynamo action driven by rotating convection in a planar geometry, exploring a subcritical region of parameter space (below the non-magnetic convective onset) where dynamically-significant magnetic fields can be sustained for multiple magnetic diffusion times. These are therefore likely examples of subcritical dynamo action. In previous studies modelling this system, subcritical dynamo action was observed for either a fraction of a magnetic diffusion time \cite{StP93} or at a single Ekman number \cite{Stel04}. Here we have investigated the subcritical range for various Ekman numbers (smaller than $3.16\times10^{-4}$) and magnetic Prandtl numbers (greater than, or equal to, unity) to determine the conditions required for this behaviour. It is worth noting that in our dynamo simulations, where the sustained magnetic field has a marked effect on the convective flow, the magnetic energy is always larger than the kinetic energy in the saturated phase, while the Elsasser number (obtained either from the standard definition based on the box-size ohmic timescale or from the modified definition based on the dynamical timescale, \S\ref{sec:def}) is always smaller than 1.  The magnetic field visibly affects the convection when the standard Elsasser number based on the mean magnetic field, $\overline{\Lambda}$, becomes order $\Ek^{1/3}$, in agreement with results from linear magnetoconvection analysis \citep{Chandra61,Elt72}.
We find that the behaviour of the nonlinear dynamo is different at moderate Ekman numbers where $\Ek\geq5\times10^{-6}$ and small Ekman numbers where $\Ek=5\times10^{-7}$. 

At moderate Ekman numbers, the flow in the saturated phase of the dynamo clearly differs from the flow in the kinematic phase: it still takes the form of columnar cells aligned with the rotation axis, but with a larger horizontal lengthscale and a localised pattern. In this regime, Lorentz and Coriolis forces are of equal importance. We found that this magnetically-modified convection generates a coherent mean magnetic field according to a dynamo process similar to the two-scale mechanism of \citet{Child72}, which operates in the kinematic phase from the small-scale convection. This dynamo mechanism thus requires a sufficient scale separation between the box size and the convective roll size.
The solutions first obtained in the supercritical regime near the onset are then tracked into the subcritical regime by gradually decreasing the Rayleigh number. Some of the subcritical branches extend down to $\Ra=0.87\Ra_c$, with the dynamo and convection collapsing for smaller $\Ra$.
We identified two necessary conditions for subcritical dynamos: (i) the magnetic Reynolds number must be larger than approximately $70-100$ to produce a coherent mean magnetic field; (ii)  $\overline{\Lambda}$ must be of order $\Ek^{1/3}$, so that the Lorentz and Coriolis forces are of equal importance at the convective scale. For a given ratio $\Ra/\Ra_c$, these two conditions are more easily met at lower $\Ek$ and larger $\Pm$, so small $\Ek$ and large $\Pm$ favour subcriticality. However, larger values of $\Pm$, and hence $\Rm$, are also associated with larger fluctuations in the velocity, which can push the system away from the dynamo branch and towards the trivial state. We therefore find optimal values of $\Ek$ and $\Pm$ for the subcritical range, which are approximately $\Ek=10^{-5}$ and $\Pm=2.5$. 

At small Ekman numbers, we obtain an intermittent dynamo in the saturated phase due to the emergence of a (near) box-size convection mode. This large-scale mode is produced in the regime where Lorentz and Coriolis forces are of equal importance when $\overline{\Lambda}\gg\mathcal{O}(\Ek^{1/3})$, a situation that arises for $\Ek=5\times10^{-7}$ in our simulations with $\Pm=1$. The large-scale mode is unable to sustain the large-scale dynamo, unlike the magnetically-modified convection at larger $\Ek$, and so this leads to cycles of growth and decay of the magnetic field. Surprisingly, despite this intermittent behaviour it appears that we are able to sustain dynamo action below the onset of convection, albeit only just below $\Ra_c$. In the subcritical cases, we still observe intermittent behaviour although it should be noted that these cases have not been run for long (approximately $10\%$ of a thermal diffusion time), so further study of this regime is needed.

Few examples of numerical dynamo simulations producing coherent large-scale magnetic fields exist in rapidly-rotating planar convection, except for the dynamos operating near the convective onset. 
Another example of an intermittent dynamo in this system is the large-scale-vortex dynamo that occurs for moderate supercritical values of the Rayleigh number (typically $\Ra>3\Ra_c$) \citep{Guervilly2017a}. This dynamo relies on the presence of the large-scale vortices that form spontaneously in non-magnetic rotating convection when the small-scale convective flow is both sufficiently turbulent (hence the condition that $\Ra>3\Ra_c$) and anisotropic due to the rapid rotation \citep{Jul12,Guer14,Favier2014}. The large-scale vortices play a crucial part in the dynamo mechanism, but tend to be disrupted when the magnetic field reaches a critical strength, leading to an intermittent behaviour. The key difference between the near-onset intermittent dynamo described here and the large-scale-vortex dynamo is that the large-scale mode of the near-onset dynamo is magnetically driven and is unable to maintain dynamo action, while the large-scale vortices at larger $\Ra$ are hydrodynamically driven and play an active role in the magnetic field generation. 
A stable (i.e. not intermittent) large-scale dynamo also exists for similar parameters as the large-scale-vortex dynamo \citep{Masada2014,Bush18}. This dynamo also requires the presence of the large-scale vortices during the kinematic phase, but the dynamo then persists without the regeneration of the large-scale vortices. However, it only works when vertical field boundary conditions are used \citep{Bush18} or when convectively stable layers are included above or below the convective layer \citep{Kapyla2013}, which permit the magnetic flux to escape from the convective domain. Dynamo simulations using perfectly conducting magnetic boundary conditions outside of the parameter windows where near-onset dynamos and large-scale-vortex dynamos operate produce only small-scale magnetic fields with no coherent large-scale fields \citep[e.g.][]{Cattaneo2006,Til12,Favier2013}. 
It is plausible that stable large-scale dynamos (which could display subcritical behaviour) exist at low Ekman numbers near the convective onset for magnetic Prandtl numbers smaller than unity, such that the Elsasser number remains sufficiently small (of order $\Ek^{1/3}$) to avoid the emergence of the magnetically-driven large-scale convection mode. We will explore this small-$\Pm$ pathway to large-scale dynamo action at low Ekman numbers in a forthcoming study.

\acknowledgements{R.G.C was supported by STFC studentship 1949517 under project ST/R50497X/1. C.G was supported by the UK Natural Environment Research Council under grant NE/M017893/1. This research made use of the Rocket High Performance Computing service at Newcastle University and the DiRAC Data Intensive service at Leicester, operated by the University of Leicester IT Services, which forms part of the STFC DiRAC HPC Facility (\href{www.dirac.ac.uk}{www.dirac.ac.uk}). The DiRAC equipment was funded by BEIS capital funding via STFC capital grants ST/K000373/1 and ST/R002363/1 and STFC DiRAC Operations grant ST/R001014/1. DiRAC is part of the National e-Infrastructure. We thank the referee for suggestions that have improved the manuscript.}


\bibliography{RGC_Subcritical_Dynamos_Final}


\end{document}